\documentclass[AMA,STIX1COL]{WileyNJD-v2}
\usepackage{bm}
\usepackage{algorithm}  
\usepackage{algorithmicx}
\usepackage{algpseudocode}  
\articletype{Article Type}%

\begin{document}

\title{Robust nonparametric integrative analysis to decipher heterogeneity and commonality across subgroups using sparse boosting}

\author[1]{Zihan Li}
\author[1]{Ziye Luo}
\author[1]{Yifan Sun}

\authormark{Zihan Li \textsc{et al}}

\address[1]{\orgdiv{Center of Applied Statistics, School of Statistics}, \orgname{Renmin University of China}, \orgaddress{\state{Beijing}, \country{China}}}

\corres{*Yifan Sun, No. 59 Zhongguancun Street, Haidian District, Beijing, 100872 \\\email{sunyifan@ruc.edu.cn}}

\abstract[Summary]{In many biomedical problems, data are often heterogeneous, with samples spanning multiple patient subgroups, where different subgroups may have different disease subtypes, stages, or other medical contexts. These subgroups may be related, but they are also expected to have differences with respect to the underlying biology. The heterogeneous data presents a precious opportunity to explore the heterogeneities and commonalities between related subgroups. Unfortunately, effective statistical analysis methods are still lacking. Recently, several novel methods based on integrative analysis have been proposed to tackle this challenging problem. Despite promising results, the existing studies are still limited by ignoring data contamination and making strict assumptions of linear effects of covariates on response. As such, we develop a robust nonparametric integrative analysis approach to identify heterogeneity and commonality, as well as select important covariates and estimate covariate effects. Possible data contamination is accommodated by adopting the Cauchy loss function, and a nonparametric model is built to accommodate nonlinear effects. The proposed approach is based on a sparse boosting technique. The advantages of the proposed approach are demonstrated in extensive simulations. The analysis of The Cancer Genome Atlas data on glioblastoma multiforme and lung adenocarcinoma shows that the proposed approach makes biologically meaningful findings with satisfactory prediction.}


\maketitle

\section{Introduction}\label{sec1}
Detection and estimation of the molecular markers associated with disease outcomes or phenotypes is one of the core tasks in biomedicine. Regression models have been extensively used to link biomarkers to a phenotypic trait. Traditionally, all samples are expected to follow the same regression model. However, in the biomedical field, the classical assumption does not always hold. The samples often span multiple subgroups, where subgroups represent, for example, disease subtypes, stages, and demographic characteristics. Different subgroups may vary in underlying etiology and pathogenesis \citealp{Curis2012The, Guinney2015The}, and therefore have distinct relationships between observed markers and a phenotype of interest. However, because all samples in subgroups have the same disease, it is reasonable to expect that they share some commonalities in the trait-marker association. To detect the important marker more accurately and understand diseases more comprehensively, it is crucial to explore the underlying heterogeneity and unveil the commonality across different subgroups concerning the trait-marker association. 

Meta-analysis is the most straightforward method, under which each subgroup is initially analyzed separately, and then summary statistics are pooled across subgroups \citealp{Guerra2009book, Yang2010A}. However, the data in biomedicine often have "large $p$, small $n$" characteristics, with the number of covariates $p$ much larger than the sample size $n$. For example, a typical cancer microarray study profiles $p\sim 10^{3-4}$ genes on $n\sim 10^{1-3}$ samples. The high-dimensionality of data makes the estimation in each subgroup challenging and often leads to unsatisfactory results for each subgroup and the overall meta-analysis \citealp{Huang2012Biostat, Huang2016JASSA}. An alternative is to conduct an integrative analysis. In contrast to meta-analysis, the integrative analysis does not develop a separate model for each subgroup/dataset; instead, it builds a joint model to analyze multiple subgroups/datasets simultaneously. Recent studies have shown that the integrative analysis often outperforms meta-analysis with more accurate marker identification and coefficient estimation \citealp{huang2017promoting1, Li2019Penalization}. Most integrative analyses focus on developing novel modeling methods to promote the accuracy of identification, estimation, and prediction. Recently, a few integrative analysis approaches have been paying special attention to the heterogeneity/commonality identification across subgroups/datasets. For example, fused lasso approach \cite[]{Tang2016Fused} and joint lasso approach \cite[]{DONDELINGER2020The} apply a fusion type penalty to encourage agreement between subgroup-specific coefficients. One of the authors and collaborators develops novel integrative analysis methods to automatically identify the commonality and difference across multiple subgroups/datasets with respect to regression coefficients of the group of covariates by using penalization \cite[]{Sun2019Identification} and boosting \cite[]{Sun2020An} techniques, respectively.

\begin{figure}
\centerline{\includegraphics[width=0.9\textwidth]{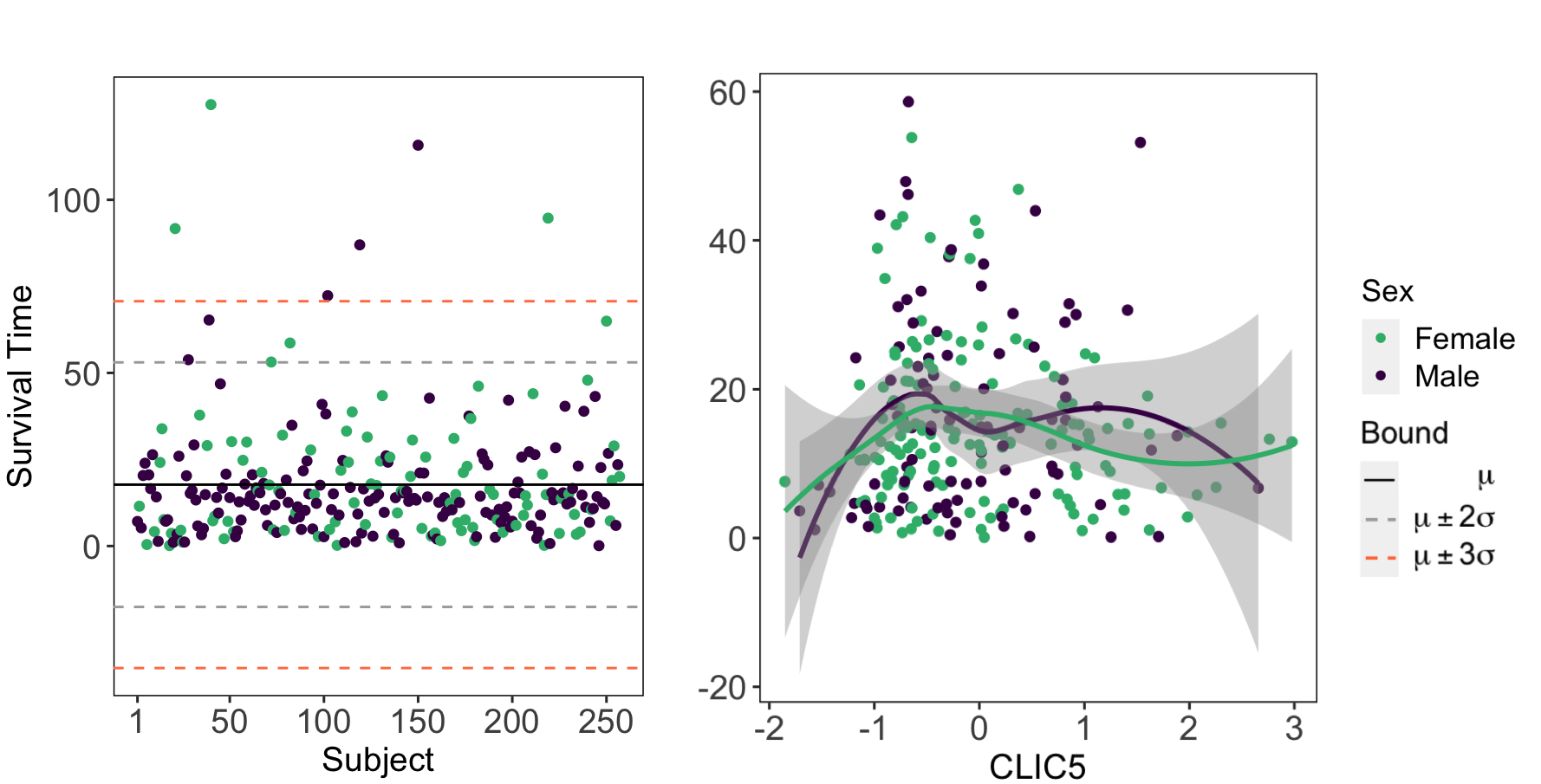}}
\caption{Analysis of TCGA GBM data. Left: scatter plot of survival time. Right: survival time against {\color{black}CLIC5} gene expression, and 
regression curve with boostrap-based $95\%$ confidence intervals (shaded area). \label{fig:scatter}}
\end{figure}

The above integrative analysis approach, although quite successful, still has limitations. First, most if not all of the existing studies lack robust properties, which are essential to accommodating outliers/contamination in the response variables. In biomedical studies, response with outliers/contamination is often encountered and can be responsible for extremely large/small response values, errors in data recording, and others \cite[]{Osborne2004The}. Specifically, consider The Cancer Genome Atlas (TCGA) glioblastoma multiforme (GBM) data analyzed in the study (for more details, please refer to Section 4.1). There are 211 deaths during follow-up, with survival time ranging from 0 to 127.5 months (median: 13.96 months). 
The left panel of Figure \ref{fig:scatter} shows the scatterplot of survival times, as well as the mean and three times standard deviation. It can be observed that six subjects are out of the three-standard-deviation range, suggesting that the data have some outliers with respect to response. The nonrobust approaches cannot effectively accommodate the outliers/contamination in response and may lead to biased estimation, false marker identification, and inference, suggesting that robust analysis is urgently needed. Robust analysis approaches have been extensively developed to low-dimensional data and shown to have clear advantages over nonrobust approaches. For the high-dimensional data, there are some robust analysis approaches, but the development is still much limited, especially for the multiple subgroups/datasets \cite[]{Filzmoser2021}. The second limitation of the existing approaches is that they usually assume that the covariates have linear effects on the response. However, recent biomedical studies have shown that nonlinear covariate effects are present in many cases \cite[]{Wu2019}, which is confirmed by a preliminary examination of the GBM data. In the right panel of Figure \ref{fig:scatter}, we present the plot of survival time (response variable) against the expressions of gene {\color{black}CLIC5}. The purple and green colors correspond to females and males, respectively, and the lines are generated by using kernel regression. It is clear that the two curves have significantly nonlinear trends, and the trends are different. Hence, it is necessary to relax the linear effect assumptions in modeling to accommodate the possible nonlinear effects of covariates on response. 

Motivated by the aforementioned limitations in current studies, we develop a robust nonparametric integrative analysis approach to identify heterogeneity and commonality in markers' associations with responses, as well as select important covariates, and estimate effects, in prognosis studies with high-dimensional molecular measurements. The proposed approach has some connections with but, more importantly, significantly advances from the existing ones. First, it adopts a Cauchy (aka Lorentzian) loss function (CLF) to accommodate outliers/contamination in response. CLF was first proposed in the computer vision field \cite[]{Black1996The} and has recently been applied to general supervised and unsupervised learning tasks for high-dimensional data \citealp{Xuelong2018Robust, Barron2019CVF}. Compared to alternative robust approaches, including the popular quantile approach and Huber's approach, CLF's approach is computationally simpler, which is critical to high-dimensional data. In addition, most current robust methods are developed for a single subgroup/dataset and have relatively limited developments for multiple subgroups/datasets. Second, a nonparametric additive model is assumed to accommodate the possible nonlinear effects of markers on response. Compared to the classical linear model, it offers a higher degree of flexibility in the form of covariate effect and can model the nonlinear covariate effects. The proposed approach is based on sparse boosting \cite[]{Buhlmann2006JMLR}, which demonstrates competitive performances in high-dimensional data analysis compared to penalization and other techniques \citealp{Yue2018Sparse, Wu2019Robust}. In all, this study is warranted by providing a practically useful new approach for exploring heterogeneity and commonality across multiple high-dimensional datasets.

\section{Methods}
\subsection{Data and model settings}
Suppose we have $M$ independent datasets. Each dataset may represent a different stage, subtype, or subpopulation of the same disease. The sample size for dataset $m$ is $n^m$, $m=1,\ldots,M$, and the total sample size is $N=\sum_{m=1}^M n^m$. In dataset $m$, denote 
$Y^{m}=(y_1^m,\ldots, y_{n^m}^m)^\top$ as the response vector and $\bm X^{m}=[(X_1^{m})^\top,\dots, (X_{n^m}^{m})^\top]^\top \in \mathbb{R}^{n^m\times p}$ as measurement matrix for $p$ covariates (e.g., gene expression, SNPs, proteomics, and so on) with $X_i^{m}=(X_{i1}^{m},\ldots,X_{ip}^{m})$. Assume that the same set of covariates are measured in $M$ datasets. Consider the following nonparametric additive model: 
\begin{equation}
\label{eq:model}
y_i^{m}\sim g \Big(\sum_{j=1}^p f_j^m(X_{ij}^m)\Big),
\end{equation}
where $g(\cdot)$ is the known link function, and $f_j^{m}(\cdot)$'s are unknown smooth functions. For identification purposes, we assume that all additive components $f_j^m$ are centered, i.e., $\sum_{i=1}^{n^m} f_j^m(X_{ij}^m)=0$ for $j=1,\ldots,p$ and $m=1,\ldots,M$. Here, the data have been normalized so that there is no intercept in the model (\ref{eq:model}). If $f_j^{m_1}(\cdot)=f_j^{m_2}(\cdot)$, then covariate $j$ represents a commonality shared by the datasets $m_1$ and $m_2$. Otherwise, it represents a heterogeneity. 

We expand each component $f_j^{m}(\cdot)$ \[f_j^{m}(\cdot)=\sum_{k=1}^K \beta_{jk}^{m}\phi_{k}(\cdot)\equiv \phi(\cdot) \beta_j^{m}\]
using basis functions $\phi(\cdot)=(\phi_1(\cdot),\ldots,\phi_K(\cdot))$, where $\beta_j^{(m)}=(\beta_{j1}^{(m)},\ldots, \beta_{jK}^{(m)})^\top$ is the unknown coefficient vector. 
To satisfy the identification restriction, the basis functions are constrained to have mean values of zero, i.e., $\sum_{i=1}^{n^m}\phi(X_{ij}^m)=0$ for $j=1,\ldots,p$ and $m=1,\ldots,M$. In the numerical study, we adopt the normalized B-spline basis function, which has been widely used in published studies \citealp{Huang2010Variable, Li2020Semi}, and note that other basis functions are also be applicable. 
Then model (\ref{eq:model}) can be rewritten as 
\begin{equation}
\label{eq:model2}
y_i^{m}\sim g\left(\sum_{j=1}^p \phi(X_{ij}^{m})\beta_j^{m}\right).
\end{equation}
Because $f_j^{m_1}(\cdot)=f_j^{m_2}(\cdot)$ is equivalent to $\beta_j^{m_1}=\beta_j^{m_2}$, we can distinguish the heterogeneity and commonality across multiple datasets according to $\beta_j^m$'s. Specifically, for covariate $j$, if $\beta_j^{m_1}=\beta_j^{m_2}$, then it represents a commonality shared by two datasets or a heterogeneity.

\subsection{Robust identification and estimation by sparse boosting}
We adopt the sparse boosting method to select important covariates, estimate the unknown effects, and more importantly, identify the heterogeneity and commonality in covariate effects across multiple datasets. Boosting methods have become very popular for high-dimensional data problems because of their low computational cost and powerful performance \citealp{Buhlmann2007Boosting, Buhlmann2010Boosting}. Although standard boosting, e.g. $L_2$Boosting \cite[]{Buhlmann2003Boosting}, are demonstrated to be effective methods in constructing a sparse model, they still tend to select many covariates irrelevant to the response. To this end, sparse boosting is developed to tackle this problem \cite[]{Buhlmann2006JMLR}, which can lead to sparser models by imposing a penalty of model complexity. The sparse boosting is originally designed to a single dataset, and has been generalized to multiple datasets \cite[]{Huang2017SIM}. Recently, Sun et al \cite[]{Sun2020An} proposed a new sparse boosting method, namely CD-Boost, to address the heterogeneity/commonality problem. The model they considered, however, is a parametric one and lacks robust properties. As shown in Figure \ref{fig:scatter}, the nonlinear covariate effects and outliers in response are not rare in biomedical data. Therefore, a modified boosting algorithm, called robust nonparametric integrative (RNP-Int) algorithm, is developed, which is summarized in Algorithm 1. 

The main difference/improvement of the proposed method is the consideration on robustness. To accommodate outliers/contamination in the response, we adopt the Cauchy loss function (CLF), and its definition is shown as follows 
\begin{equation*}
\rho_c(x)=\log\left(1+(x/c)^2\right) 
\end{equation*}
with influence function 
\begin{equation*}
\psi_c(x)\equiv\frac{\partial \rho_c(x)}{\partial x}=\frac{2x}{x^2+c^2},
\end{equation*}
which is used to measure the effect of changing the point of the sample on the value of the parameter estimation. Here, $c$ is a constant. We demonstrate CLF function and its influence function in Figure \ref{fig:CLF} (Appendix C). Compared to the least-squares loss with $\rho(x)=x^2$, CLF's influence function has an upper bound. Although the least-absolute deviation loss with $\rho(x)=|x|$ as well as the Huber loss \cite{} can alleviate the effect of the large error, their influence functions have no cutoff \cite[]{He2000Breakdown}. In contrast, CLF's influence function has an upper bound and its values tend to reduce to zero with the increase of the error. Therefore, it can alleviate the influence of a single sample, especially a sample with a large noise, on estimating the residuals. As a result, CLF has less dependence on the distribution of the noise and is more robust to the noise. When $g$ is a linear regression model, for example, we use CLF to penalize the residual terms, which is defined as 
\begin{equation*}
\rho_c(\beta^{m})= \sum_{i=1}^{n^m} \log \left(1+\frac{\left(y_i^m-\sum_{j=1}^p \phi(X_{ij}^{m})\beta_j^{m}\right)^2}{c^2}\right), m=1,\ldots,M,
\end{equation*}
where $\beta^m=\left((\beta_1^m)^\top,\ldots,(\beta_p^m)^\top\right)^\top$. In this study, we also consider right censored survival responses under accelerated failure time (AFT) model. Details on the model settings and loss functions are provided in Appendix A.

Beyond robustness, there are three main distinctions between the proposed and standard sparse boosting algorithms for multiple datasets. 
First, when determining in each iteration which covariate is to be included in the model, multiple datasets are considered jointly. Second, the increment $\gamma$ in coefficient vectors of $M$ datasets are determined simultaneously. Specifically, for covariate $j$, we divide $M$ datasets into $L_j$ disjoint groups $(G_1,\ldots,G_{L_j})$, where each group $G_l$ has an identical coefficient vector, i.e., $\beta_j^{m_1}=\beta_j^{m_2}$, for any $m_1\in G_l$ and $m_2\in G_l$, and different groups have distinct coefficient vectors, i.e., $\beta_j^{m_1}\neq \beta_j^{m_2}$, for any $m_1\in G_{l_1}$ and $m_2\in G_{l_2}$ with $l_1\neq l_2$. Some datasets in $G_l (l=1,\ldots, L_j)$ are allowed to have identical increments in coefficient vectors to maintain commonality between these datasets. {\color{black} More precisely, for each unempty subset $G$ of $G_l$ $(l=1,\ldots, L_j)$,  we compute the optimal common increment $\hat{\gamma}_G$ in coefficient vectors of datasets in $G$ by minimizing the sum of CLFs (\ref{eq:obj1}). Then, we select the subset $G^j$ and the corresponding increment $\gamma^j$ in their coefficients vectors which minimizes the objection function $\mathcal{F}(G,\hat{\gamma}_G; j)$ (\ref{eq:obj}) based on enumeration method. 
The objection function $\mathcal{F}(G,\hat{\gamma}_G; j)$ has three components. The first component is the loss function, which is the sum of CLFs of $M$ datasets.  
The second component explicitly penalizes model complexity and hence can promote sparsity. In the literature, there are multiple choices for model complexity measures, for example, Akaike information criterion (AIC), Bayesian information criterion (BIC), cross-validation, and minimum description length (MDL). Here we use a BIC type measure as an example, which has been employed in published studies. The second penalty explicitly quantifies difference across multiple datasets and hence can encourage the commonality of coefficient vectors, or equivalently, additive components, across multiple datasets. This penalty was firstly proposed in \cite[]{Sun2020An}, and the introduction of this penalty is another distinction of the proposed method. } It is proportional to the number of covariates with different coefficient vectors in two distinct datasets, and takes value in $[0,1]$. It is minimized when all coefficient vectors are the same across $M$ datasets, that is, $\beta_j^1 = \ldots = \beta_j^M$ for $j=1,\ldots, p$, and is maximized when all coefficient vectors are different. $\lambda$ is the tuning parameter, which controls the strength of penalization. On the one hand, when $\lambda=0$, the penalization disappears, then the proposed method degenerates into the standard sparse boosting algorithm for multiple datasets \cite[]{Huang2017SIM}; on the other hand, when $\lambda=\infty$, the proposed method compels $M$ datasets to have identical $p$ coefficient vectors, i.e., $\beta_j^1=\ldots=\beta_j^M$, for $j=1,\ldots, p$.  

\begin{algorithm}
\caption{RNP-Int}\label{alg1}
\begin{algorithmic}[1]
\State \textbf{Initialization}. Denote $\bm \beta=(\beta^{1},\ldots,\beta^{M})$, and $\bm\beta^{(t)}=(\beta^{1(t)},\ldots, \beta^{M(t)})$ as the estimate of $\bm \beta$ in the $t$th iteration. Set $\bm\beta^{(1)}=\bm 0$. 
\For {$t=1,\ldots, T$} 
\For {$j=1,\ldots, p$} 
\State Denote $L_j$ as the number of unique vectors in $(\beta_j^{1(t)},\ldots,\beta_j^{M(t)})$. Let $(G_1,\ldots, G_{L_j})$ be a mutually exclusive partition of $\{1,\ldots,M\}$, such that $\beta_j^{m(t)}=\alpha_l$ for all $l\in G_l$, where $\alpha_l$ is the common value for the $\beta_j^{m(t)}$'s from group $G_l$. 
\State Define $\mathcal{G}=\{G|G\subset G_l, l=1,\ldots,L_j\}\setminus \emptyset$ and $\gamma=(\gamma_1,\ldots,\gamma_K)^\top$. 
{\color{black}
\State Compute 
\begin{equation}
\label{eq:obj1}
\hat{\gamma}_G=\text{argmin}_{\gamma}\sum_{m\in G}\rho_c\left(\beta^{m(t)}+\bm A_j\gamma\right), \ \ G\in \mathcal{G},
\end{equation}
where $\rho_c(\cdot)$ is Cauchy loss function (details below). Here $\bm A_j=(1_{(j-1)K+1},\ldots, 1_{jK})$ with $1_q$ is the length-$pK$ vector with the $q$th component equal to 1 and all others equal to 0. 

\State To determine optimal subset $G^j$ and increment of coefficient vectors $\gamma^j$, compute 
\begin{equation}
\begin{aligned} 
\label{eq:obj}
(G^j,\gamma^j)= & \text{argmin}_{\{(G,\hat{\gamma}_G)|G\in\mathcal{G}\}} \mathcal{F}(G, \hat{\gamma}_G;j)\equiv \sum_{m=1}^M \left[\rho_c\left(\beta^{m(t)}+\chi(m\in G)\bm A_j\hat{\gamma}_G\right)+\frac{\log n^m}{n^m}\sum_{s=1}^p \chi\left(\beta_s^{m(t)}+\chi(m\in G)\delta_{sj}\hat{\gamma}_G\neq 0\right)\right]\\
& +\lambda pen_c\left(\bm \beta^{(t)}+\sum_{m\in G}\bm{1}_{,m}\bm A_j\hat{\gamma}_G\right)
\end{aligned}
\end{equation} 
where $\chi(\cdot)$ is the indicator function, and $\bm 1_{,m}$ is a $pK\times M$ matrix with $m$th column being 1s and other being zeros. Here $pen_c(\bm\beta)=1-\sum_{m_1<m_2}\sum_{s=1}^p\frac{\chi\left(\beta_s^{m_1}=\beta_s^{m_2}\right)}{\tbinom{M}{2}p}$, and $\lambda$ is the tuning parameter. }
\EndFor
\State Determine the optimal $j$ as $\hat j=\text{argmin}_{1\leq j \leq p}\mathcal{F}(G^j, \gamma^j;j)$. 
\State Update {\color{black}$\bm\beta^{(t+1)}=\bm\beta^{(t)}+v \sum_{m\in G^{\hat j}}\bm{1}_{,m} \bm A_{\hat{j}}\gamma^{\hat{j}}$}, where $v$ is the step size. It has been suggested that the choice of $v$ is not critical as long as it is small\cite[]{Buhlmann2006JMLR}, and we set $v=0.1$ following the published studies. 
\State Compute \[S(t)=\sum_{m=1}^M \left[\rho_c(\beta^{m(t+1)})+\frac{\log n^m}{n^m}\sum_{s=1}^p \chi\left(\beta_s^{m(t+1)}\neq 0\right)+\lambda pen_c(\bm\beta^{(t+1)})\right]\]  
\EndFor
\State Select the optimal stopping iteration by $t^\star=\text{argmin}_{t}S(t)$. 
\State $\bm\beta^{(t^\star)}=(\beta^{1(t^\star)}, \ldots, \beta^{M(t^\star)})$ is the final estimator of $\bm\beta$. The strong learner for dataset $m$ is $L_i^m=g(\sum_{j=1}^p\phi(X_{ij}^m) \beta_j^{m(t^\star)})$ ($i=1,\ldots, n^m$). 
 \end{algorithmic}
\end{algorithm}

\noindent{\bf Tuning parameter.} The proposed method involves two tuning parameters $c$ and $\lambda$. For $c$, Li et al \cite[]{Xuelong2018Robust} suggest taking some small values (for example, 0.1, 0.5, 1, and 2). In our numeric study, we find that the results are not sensitive to $c$ and set $c=1$ to reduce the computation burden. In practice, one may need to examine multiple $c$ values and select the optimal value using data-driven methods. $\lambda$ controls the commonality among components in multiple datasets. We search for the optimal $\lambda$ value using five-fold cross-validation (CV). Specifically, each dataset is partitioned randomly into five disjoint subsets with equal sizes. We apply a one-dimensional grid search for $\lambda$ with the search range as $\{2^{-6}, 2^{-5},\ldots, 2^{-1}\}$. 

\noindent{\bf Realization.} To facilitate data analysis within and beyond this study, we have developed a mathematics code implementing the proposed method and made it publicly available at HTTPS: //github.com/sunyifan1984/RNP-Int. Further, we provide a demo for two datasets with continuous and survival outcomes, respectively.  The proposed method is computationally affordable. For example, for a simulation example with $m=3$, $n^m=100$ ($m=1,2,3$), and $p=200$, overall analysis can be accomplished within 15 min using a laptop with standard configurations. The computation time spent on tuning parameters can be greatly reduced by parallel computing.

\section{Application to simulation data}\label{sec3}
\subsection{Data generation procedure}
This section aims to assess the performance of the proposed method through a simulation study. Suppose there are $M=3$ independent datasets with, for simplicity, equal sample size $n^1=n^2=n^3=100$. $p=200$ covariates are generated from a multivariate normal distribution with zero means and unit variances. We consider an autoregressive correlation structure where covariates $j$ and $k$ have a correlation coefficient $0.5^{|j-k|}$. In each dataset, there are 3-6 covariates with nonzero additive components. Concerning the heterogeneity/commonality across three datasets, we consider three scenarios. Firstly, 
one covariate has identical nonzero components in three datasets, three covariates have the same nonzero components in two of the three datasets, and one covariate has different nonzero components in different datasets. Secondly, all datasets have the same set of important covariates and also the same components, and finally, none of the datasets share the same components. The nonzero additive components in three scenarios are listed below. 
\begin{itemize}
\item Scenario 1 for components: \\
$f_1^1(x)=x$, $f_3^1(x)=3\sin(x)$, $f_4^1(x)=6/(1+e^{-2x})-3$, $f_6^1(x)=2x$;\\
$f_1^2(x)=x$, $f_2^2(x)=x^2-1$, 
$f_5^2(x)=3e^{-x^2/4}-\sqrt{6}$;\\
$f_1^3(x)=x$, $f_2^3(x)=x^2-1$, $f_3^1(x)=3\sin(x)$, $f_5^3(x)=3e^{-x^2/4}-\sqrt{6}$, $f_6^3(x)=-2x$;
\item Scenario 2 for components: \\
$f_1^m(x)=x$, $f_2^m(x)=x^2-1$, $f_3^m(x)=2\sin(x)$, $f_4^m(x)=6/(1+e^{-2x})-3$, $f_5^m(x)=3e^{-x^2/4}-\sqrt{6}$, $f_6^m(x)=2x$ for $m=1,2,3$;
\item Scenario 3 for components: \\
$f_1^1(x)=x$, $f_2^1(x)=x^2-1$, $f_3^1(x)=2\sin(x)$, $f_4^1(x)=6/(1+e^{-2x})-3$, $f_5^1(x)=3e^{-x^2/4}-\sqrt{6}$, $f_6^1(x)=2x$;\\
$f_2^2(x)=2(x^2-1)$, $f_3^2(x)=4\sin(x)$, $f_5^2(x)=-3e^{-x^2/4}+\sqrt{6}$, $f_6^2(x)=4x$;\\
$f_1^3(x)=2x$,  $f_3^1(x)=-\sin(x)$, $f_4^3(x)=-6/(1+e^{-2x})+3$, $f_6^3(x)=-2x$.
\end{itemize}
The rest of the components are zeros. In the aforementioned scenarios, each dataset has some nonlinear components. We also examine the scenario where all nonzero components degenerate into linear functions. Specifically, 
\begin{itemize}
\item Scenario 4 for components: \\ 
$f_1^1(x)=f_3^1(x)=f_4^1(x)=x$, $f_5^1(x)=f_6^1(x)=2x$, $f_9^1(x)=1.5x$; \\
$f_1^2(x)=f_2^2(x)=f_4^2(x)=x$, $f_7^2(x)=-x$, $f_8^2(x)=2x$, $f_9^2(x)=-1.5x$; \\
$f_1^3(x)=f_2^3(x)=f_3^3(x)=x$, $f_6^3(x)=-2x$, $f_7^3(x)=2x$, $f_9^3(x)=3x$, \\
\end{itemize}
where each dataset has six important covariates, and share one common important covariate and also the identical component. 

We consider continuous data under the additive model and right-censored survival data under the AFT model with the following error distributions: (1) Error 1: N(0,1), (2) Error 2: 70\%N(0,1)+30\%Cauchy(0,1), and (3) Error 3: Cauchy(0,1). With censored survival data, the censoring times are generated independently from uniform distributions. The censoring distributions are adjusted so that the overall censoring rate is about $20\%$. In both simulation and real data analysis, we adopt a cubic $B$-spline basis with $6$ equally spaces inner knots for each $f_{j}^m$ and without intercept; thus the dimension $K$ of the B-spline basis is 6. 

\subsection{Alternative approaches}
Besides the proposed robust nonparametric integrative (RNP-Int) approach, we also consider the following alternatives for comparison. The nonrobust nonparametric integrative (referred to as the "NRNP-Int") approach has the same modeling framework as the proposed one. However, it adopts nonrobust loss, i.e., least-squares loss, in estimation, selection procedure, and stopping criterion. The robust parametric integrative (referred to as the "RP-Int") approach is the same as the proposed, except that all covariate effects are assumed to be linear. The nonrobust parametric integrative (referred to as "NRP-Int") is the nonrobust counterpart of RP-Int. The robust nonparametric meta (referred to as the "RNP-Meta") approach applies sparse boosting under robust nonparametric modeling to each dataset separately, and then results are combined across datasets. The robust nonparametric pool (referred to as the "RNP-Pool") approach combines all datasets and applies robust nonparametric sparse boosting. Details on the alternatives RNP-Meta and RNP-Pool are provided in Appendix B. We acknowledge that other integrative analysis approaches can be adopted to analyze the simulated data. The above five alternatives are the most direct competitors as they follow a similar framework as the proposed approach. Specifically, by comparing with NRNP-Int and  NRP-Int as well, the robustness of the proposed approach can be directly established. The advantages of nonparametric modeling of the proposed approach can be demonstrated by comparing it with RP-Int. The significance of adopting integrative analysis to identify heterogeneity/commonality across datasets is investigated by comparing with RNP-Meta and RNP-Pool.  

For each training data, we also generate independent test data with $n^1=n^2=n^3=50$ in the same manner. The train datasets are used to train models, and test datasets are used to measure the prediction accuracy of estimated models. For all six methods, the maximum number of boosting iterations is limited to $T=300$. 

\subsection{Performance comparison}

\begin{center}
\begin{table}[t]
   \centering
   \caption{
   Simulation Scenarios 1-2 for data with a continuous outcome. In each cell, mean (standard deviation).} 
   \begin{tabular}{llcccccc} 
      \toprule
      &   & \multicolumn{2}{c}{\bf Commonality identification} & \multicolumn{2}{c}{\bf Variable selection}  \\
      \cmidrule(r){3-4} \cmidrule(l){5-6} 
        & {\bf Methods} & \bf {TP-ind} &{\bf FP-ind }&  {\bf TP-var}  & {\bf FP-var} & {\bf RMISE} & {\bf MAE} \\
      \midrule
       & & & & \bf{ Scenario 1} \\
       Error 1 & RNP-Int & 6.75 (0.55)& 1.60 (0.82)& 11.20 (0.41) & {0.15 (0.37) } & 2.01 (0.20) & 2.87 (0.20)\\
       N(0,1) 
        & NRNP-Int & 6.70 (0.66)& {0.30 (0.73)}  & {11.90 (0.31) } & 0.60 (0.68) & {1.80 (0.20) }&{2.78 (0.17)} \\
       & RP-Int & 6.60 (0.68) & 4.20 (0.70) & 8.90 (0.31) & 0.16 (0.38) & 3.83 (0.80) & 4.03 (0.43) \\
       & NRP-Int  & 6.05 (1.00) & 3.60 (0.68) & 9.35 (0.59) & 7.60 (3.62) & 6.21 (1.20) & 4.45 (0.38)  \\ 
       & RNP-Meta  & 1.00 (0.00) & 0.30 (0.73) &11.70 (0.66) & 3.25 (1.55) & 3.52 (1.40) & 3.09 (0.31)  \\
       & RNP-Pool &  7.00 (0.00)& 11.00 (0.00)&  9.20 (0.41)  & 3.40 (0.82) & 10.57 (0.60) & 8.03 (0.71) \\
       \specialrule{0em}{3pt}{3pt}
       Error 2 & RNP-Int & {6.55 (0.83)} & 1.80 (0.62) & 11.10 (0.31) & 0.20 (0.41) & 2.14 (0.40) & 13.26 (6.33)  \\
       70\%$N(0,1)$ & NRNP-Int & 5.60 (1.47) & 5.20 (3.94) & 7.15 (4.11) & 17.20 (4.96) & 43.20 (7.88) & 24.66 (6.38)\\
       +30\%Cauchy(0,1) 
       & RP-Int & 6.55 (0.69) & 4.20 (0.70) & 8.90 (0.31) & 0.22 (0.38) & 3.75 (0.82) & 14.34 (6.33) \\
       & NRP-Int  & 5.80 (1.28) & 6.00 (3.11) & 6.00 (3.39) &19.65 (3.39) & 43.82 (8.12) & 23.52 (6.40) \\  
       & RNP-Meta  & 1.00 (0.00) & 0.75 (1.33) & 11.40 (1.14) & 4.75 (2.61) & 4.60 (2.60) & 13.72 (6.45)  \\
       & RNP-Pool & 7.00 (0.00) & 11.00 (0.00) & 9.10 (0.31) & 3.20 (0.62) & 10.60 (0.40) & 18.21 (6.35)  \\
       \specialrule{0em}{3pt}{3pt}
       Error 3 & RNP-Int & 6.25 (1.07)  & 2.40 (0.82) & 11.00 (0.00) & 0.70 (1.17) & 3.08 (1.00) & 13.55 (1.74) \\
       Cauchy(0,1) 
       & NRNP-Int & 5.35 (1.27) & 9.05 (2.24) & 1.75 (2.49) & 21.30 (2.75) & 196.11 (36.52) & 71.16 (11.10) \\
       & RP-Int   & 6.45 (0.76) & 4.30 (0.73) & 8.85 (0.49) & 0.50 (0.69) & 4.42 (1.24) & 14.34 (1.75) \\
       & NRP-Int & 6.05 (1.28) & 9.45 (1.70) & 1.75 (2.38) & 21.50 (3.22) & 202.38 (37.56) & 64.62 (9.69) \\
       & RNP-Meta  &  1.15 (0.59) & 2.20 (0.95) & 9.65 (1.27) & 11.15 (3.13) & 10.26 (3.20) & 14.95 (1.74)  \\
       & RNP-Pool & 7.00 (0.00) & 11.00 (0.00) & 9.30 (0.73) & 3.45 (1.10) & 11.20 (1.00)  & 17.75 (1.79) \\
       \midrule
        & & &  & \bf{Scenario 2} \\
        Error 1 & RNP-Int & 18.00 (0.00) & 0.00 (0.00) & 18.00 (0.00) & 0.00 (0.00) & 2.12 (0.20) & 2.77 (0.19)   \\
        N(0,1) 
       &NRNP-Int & 17.90 (0.45)  & 0.00 (0.00) & 18.00 (0.00) & 0.55 (0.94) & 2.29 (0.40)  & 2.79 (0.19)   \\ 
       & RP-Int   &  18.00 (0.00) & 0.00 (0.00) & 12.00 (0.00) & 0.00 (0.00) & 5.57 (0.18) & 4.60 (0.45)\\
       & NRP-Int &  15.05 (2.21) & 0.00 (0.00) & 13.00 (1.45) & 12.45 (2.95) & 10.37 (0.97) & 5.24 (0.41)\\
       & RNP-Meta  & 0.55 (1.10) & 0.00 (0.00) & 16.30 (1.69) & 3.95 (3.20) & 6.55 (3.40) & 3.75 (0.79)   \\
       & RNP-Pool & 18.00 (0.00) & 0.00 (0.00) & 18.00 (0.00) & 0.00 (0.00) & 2.12 (0.20) & 4.13 (0.37)   \\
        \specialrule{0em}{3pt}{3pt}
         Error 2  & RNP-Int & 18.00 (0.00) & 0.00 (0.00)  & 18.00 (0.00) & 0.15 (0.67)  & 2.20 (0.40) & 13.10 (6.32) \\
        70\%$N(0,1)$  & NRNP-Int &  15.70 (1.63) & 0.00 (0.00) & 11.45 (5.37) & 17.20 (4.42) & 45.44 (8.36) & 24.84 (6.41) \\
       30\% Cauchy(0,1) & RP-Int  &  17.90 (0.45) & 0.00 (0.00) & 12.00 (0.00) & 0.15 (0.67) & 5.81 (0.42) & 14.89 (6.32)\\
       & NRP-Int & 15.85 (1.79) & 0.00 (0.00) & 9.05 (3.93) & 19.70 (2.58) & 46.80 (8.24) & 24.12 (6.44) \\
       & RNP-Meta  & 1.15 (1.69) & 0.00 (0.00) & 15.35 (2.37) & 6.95 (3.79) & 9.40 (4.60) & 14.80 (6.36) \\
       & RNP-Pool & 18.00 (0.00) & 0.00 (0.00) & 18.00 (0.00) & 0.00 (0.00) & 2.26 (0.40) & 14.44 (6.32) \\
        \specialrule{0em}{3pt}{3pt}
       Error 3 & RNP-Int & 17.90 (0.45) & 0.00 (0.00)  & 17.40 (1.23) & 0.25 (0.91) & 3.22 (1.40) &13.49 (1.75) \\
       Cauchy(0,1) & NRNP-Int & 13.40 (1.67) & 0.00 (0.00) & 3.05 (3.83) & 21.75 (3.82) & 205.14 (37.48) & 71.91 (10.82)\\
       & RP-Int   & 16.30 (1.65) & 0.00 (0.00) & 12.00 (1.26) & 0.75 (1.33) & 6.37 (0.62) & 14.74 (1.76) \\
       & NRP-Int & 15.70 (2.18) & 0.00 (0.00) & 3.00 (3.37) & 21.00 (2.08) & 211.4 (39.36) & 65.78 (9.99)\\
       & RNP-Meta  & 3.20 (1.64) & 0.00 (0.00) & 11.70 (2.00) & 13.95 (3.83) & 17.79 (4.40) & 16.46 (1.68) \\
       & RNP-Pool & 18.00 (0.00) & 0.00 (0.00) & 17.40 (2.00) & 0.15 (0.67) & 3.16 (1.20) & 14.57 (1.75)\\
 \bottomrule
   \end{tabular}
   \label{tab:sce1-2}
\end{table}
\end{center}
For the proposed approach and five alternatives, we evaluate the heterogeneity/commonality identification performance by the the true positive (TP-ind) and false positive (FP-ind). When computing the number of positives, for a covariate $j(=1,\ldots, 6$ in Scenarios 1-3, and $=1,\ldots, 9$ in Scenario 4), we count the number of pairs of distinct datasets where this covariate has the same effect, and then sum the first six (nine) covariates. Specifically, the true number of positives for Scenarios 1-4 equals {\color{black}7, 18, 0, and 8}, respectively. We adopt the true positive (TP-var) and false positive (FP-var) to evaluate the variable selection performance, where the number of positives is the number of true nonzero components. Estimation and prediction are also evaluated. Specifically, estimation is measured by the root mean integrated squared errors (RMISE), which is defined as $\sqrt{\sum_{m=1}^M\frac{1}{n^m}\sum_{j=1}^p\sum_{i=1}^{n^m}\left(\hat{f}^m_j(X_{ij}^m)-f_j^m(X_{ij}^m)\right)^2}$. {\color{black} Prediction is measured by the mean absolute error (MAE) for the continuous outcome, which is defined as $\sum_{m=1}^M \sum_{i=1}^{n^m}|\hat{y}_i^m-y_i^m|/n^m$, and C statistics (Cstat) for the survival outcome. The Cstat is the time-integrated area under the time-dependent ROC and measures the overall risk prediction quality for survival data, of which the values are between 0.5 (random prediction) and 1 (perfect prediction).} In each scenario, we calculate the average and standard deviation of the six metrics over 100 independent runs. 

For data with a continuous outcome, the summary results for Scenarios 1-2 and 3-4 are reported in Table \ref{tab:sce1-2} and \ref{tab:sce3-4}, respectively. The results for right-censored survival data are shown in Table \ref{tab:AFTsce1-2}-\ref{tab:AFTsce3-4} (Appendix C). In general, the proposed RNP-Int performs the best in the identification of heterogeneity/commonality in the majority of the scenarios. Without data contamination, RNP-Int and the nonrobust counterpart NRNP-Int have favorable performances. For example, in Scenario 1, under error 1, RNP-Int and NRNP-Int have (TP-ind, FP-ind)=(6.75, 1.60) and (6.70, 0.30), respectively, compared to (6.60, 4.20) for RP-Int, (6.05, 3.60) for NRP-Int, (1.00, 0.30) for RNP-Meta, and (7.00, 11.00) for RNP-Pool. When data have contamination, RNP-Int performs significantly better than the alternatives. For example, in Scenario 1, under error 2, RNP-Int has (TP-ind, FP-ind)=(6.55, 1.80), which is superior to the alternatives: (5.60, 5.20) for NRNP-Int, (6.55, 4.20) for RP-Int, (5.80, 6.00) for NRP-Int, (1.00, 0.75) for RNP-Meta, and (7.00, 11.00) for RNP-Pool. It is also worth noting that identification results by the robust methods generally have smaller standard deviations. Comparison with the robust parametric integrative method, RP-Int, also shows the merits of the proposed method. When some of the components are nonlinear functions (Scenarios 1-3), RP-Int identifies fewer true positives and (or) more false positives. As expected, RP-Int has superior performance when all of the components are linear functions (Scenario 4). It is noted that RNP-Int still has competitive performance even in the worst case, thus providing a safe choice in practice. RNP-Meta and RNP-Pool have the worst performance among all the methods. Specifically, RNP-Meta can hardly identify commonality. RNP-Pool assumes that all datasets follow the same model. As a result, it fails when there are large differences across datasets (Scenarios 1, 3, and 4). 

In identifying important covariates, the proposed RNP-Int again has the best or close to the best performance compared with alternatives. In general, RNP-Int can identify more true positives and fewer false positives. The advantage of RNP-Int against its nonrobust counterpart NRNP-Int becomes more pronounced when data have contamination (errors 2 and 3). When all components are linear functions (Scenario 4), the parametric integrative method, RP-Int, has good performance. However, as expected, its performance deteriorates significantly when some components become nonlinear. Across all simulation scenarios, RNP-Meta performs worse than the proposed RNP-Int. Among the robust methods, RNP-Pool has the worst performance when multiple datasets have completely different components (Scenario 3), which is as expected. Although its performance improves with the increase of the level of commonality across datasets, it is still inferior to RNP-Int when there are some differences between datasets. 

In terms of estimation and prediction, the proposed RNP-Int method, in general, has better performance. For example, in Scenario 1, under error 3, RNP-Int has smaller estimation and prediction errors (3.08, 13.55) than NRNP-Int (196.11, 71.16), RP-Int (4.42,14.34), NRP-Int (202.38, 32.59), RNP-Meta (10.26, 14.95), and RNP-Pool (11.20, 17.75). To visualize the estimation results better, we show the estimation of components under Scenario 1, error 3 (Figure \ref{fig:sim} in Appendix C). It can be observed that the proposed approach provides a more accurate estimation. For data with a survival outcome (Table \ref{tab:AFTsce1-2}-\ref{tab:AFTsce3-4} in Appendix C), the overall observed patterns are similar, where the proposed approach has a favorable performance in identification, estimation, and prediction.

{\color{black} In the aforementioned simulation, all datasets have the same number of samples. We also examine the scenario where different datasets have different sample size. Consider the case with $n^1=130$, $n^2=110$, and $n^3=60$. The other settings are the same as described above. We use Scenario 1 as a representative and present results in Table \ref{tab:unbalance1} (Appendix C) for data with a continuous outcome and Table \ref{tab:unbalance2} (Appendix C) for data with a right-censored survival outcome. The proposed RNP-Int is again observed to have favorable performance in identification of heterogeneity/commonality, variable selection, estimation, and prediction. }

\begin{table}[htbp]
   \centering
   \caption{
   Simulation Scenarios 3-4 for data with a continuous outcome. In each cell, mean (standard deviation).} 
   \begin{tabular}{llcccccc} 
      \toprule
      & & \multicolumn{2}{c}{{\bf Commonality identification}} & \multicolumn{2}{c}{{\bf Variable selection}}  \\
      \cmidrule(r){3-4} \cmidrule(l){5-6} 
      & {\bf Methods} & \bf {TP-ind} &{\bf FP-ind }&  {\bf TP-var}  & {\bf FP-var} & {\bf RMISE} & {\bf MAE} \\
      \midrule
       & & & & \bf{ Scenario 3} \\
       Error 1  
       & RNP-Int & 0.00 (0.00) & 2.35 (1.35) & 12.85 (0.93) & 0.80 (0.70) & 5.76 (2.00) & 3.78 (0.50) \\
       N(0,1) 
       & NRNP-Int & 0.00 (0.00) & 1.15 (0.59) & 13.60 (0.50) & 0.60 (0.82) & 4.35 (0.80) & 3.29 (0.22)  \\
       & RP-Int  &0.00 (0.00) & 6.90 (1.02) & 9.75 (0.44) & 0.65 (0.67) & 7.41 (0.78) & 5.39 (0.57)  \\
       & NRP-Int  &0.00 (0.00) & 6.25 (1.12) & 10.00 (0.65) & 9.75 (2.36) & 13.22 (1.58) & 6.31 (0.62)\\
       & RNP-Meta  & 0.00 (0.00) & 0.35 (0.59) & 13.65 (0.59) & 3.2 (2.14) & 4.76 (2.05) & 3.39 (0.45)    \\
       & RNP-Pool & 0.00 (0.00) & 18.00 (0.00) & 11.85 (1.60) & 3.00 (0.73) & 18.40 (1.00) & 9.97 (1.49) \\
       \specialrule{0em}{3pt}{3pt}
       Error 2 
       & RNP-Int & 0.00 (0.00) & 3.00 (2.49) & 12.50 (1.40) & 0.45 (0.69) & 6.14 (2.40) & 14.31 (6.33) \\
       70\% N(0,1)
        & NRNP-Int & 0.00 (0.00) & 8.50 (5.32) & 8.55 (4.64) & 14.55 (5.74) & 46.79 (8.48) & 25.74 (6.44) \\
       +30\%Cauchy(0,1) 
       & RP-Int   & 0.00 (0.00) & 7.40 (1.23) & 9.55 (0.76) & 0.70 (0.73) & 8.12 (1.39) & 15.74 (6.33) \\
       & NRP-Int & 0.00 (0.00) & 10.30 (3.50) & 7.15 (3.18) & 17.40 (4.38) & 47.41 (7.92) & 24.94 (6.35) \\
       & RNP-Meta  & 0.00 (0.00) & 0.95 (1.15) & 13.00 (1.17) & 7.15 (3.25) & 7.83 (3.41) & 14.38 (6.40) \\
       & RNP-Pool & 0.00 (0.00) & 18.00 (0.00) & 11.55 (1.90) & 2.85 (0.81) & 18.46 (1.22) & 20.05 (6.36)  \\
       \specialrule{0em}{3pt}{3pt}
       Error 3 
       & RNP-Int & 0.00 (0.00) & 4.30 (1.89) & 11.80 (0.70) & 1.05 (0.89) & 7.50 (1.80) & 14.55 (1.74) \\
       Cauchy(0,1) 
       & NRNP-Int & 0.00 (0.00) & 14.75 (3.63) & 2.80 (2.95) & 21.15 (2.54) & 204.07 (37.36) & 73.65 (10.81) \\
       & RP-Int   & 0.00 (0.00) & 8.30 (1.72) & 9.40 (0.82) & 1.55 (1.61) & 9.35 (1.99) & 15.82 (1.74) \\ 
       & NRP-Int &  0.00 (0.00) & 15.00 (2.77) & 2.50 (2.93) & 20.50 (1.76) & 212.2 (39.24) & 66.85 (9.58)\\
       & RNP-Meta  &  0.00 (0.00) & 2.30 (1.75) & 11.20 (1.99) & 12.10 (4.09) & 13.78 (5.80) & 15.85 (1.76)  \\
       & RNP-Pool & 0.00 (0.00) & 18.00 (0.00) & 10.60 (1.47) & 2.75 (1.45) & 19.22 (1.38) & 19.47 (1.76) \\
       \midrule
        & & &  & \bf{Scenario 4} \\
        Error 1 
       & RNP-Int & 7.30 (0.66) & 2.10 (2.49) & 16.75 (2.10) & 1.10 (1.25) & 5.78 (1.49) & 3.47 (0.84)  \\
      N(0,1)  &NRNP-Int & 6.90 (1.37) & 0.35 (0.88) & 17.95 (0.22) & 0.85 (0.81) & 3.98 (0.99) & 2.91 (0.21)  \\
       & RP-Int & 7.45 (0.69) & 0.70 (1.30) & 17.80 (0.52) & 1.05 (1.19) & 2.02 (1.02) & 2.75 (0.28)\\
       & NRP-Int & 6.80 (1.06) & 0.00 (0.00) & 18.00 (0.00) & 0.65 (0.93) & 1.58 (0.38) & 2.62 (0.15) \\
       & RNP-Meta  &  0.95 (0.22) & 0.45 (1.10) & 13.40 (1.05) & 3.35 (2.41) & 5.40 (2.28) & 3.23 (0.62) \\
       & RNP-Pool &  8.00 (0.00) & 19.00 (0.00) & 12.60 (1.67) & 5.10 (0.85) & 25.46 (1.42) & 12.34 (1.02)\\
        \specialrule{0em}{3pt}{3pt}
       Error 2 
       & RNP-Int &  6.30 (1.30) & 2.90 (2.10) & 15.90 (1.62) & 1.10 (0.97) & 6.78 (2.36) & 14.09 (6.35)\\
       70\%$N(0,1)$  
       & NRNP-Int  & 6.35 (1.39) & 7.45 (5.18) & 11.30 (5.15) & 23.55 (8.49) & 54.57 (9.12) & 26.15 (6.38)\\
       +30\% Cauchy(0,1)  & RP-Int &  6.65 (1.23) & 1.90 (1.48) & 17.15 (0.99) & 1.20 (1.06) & 3.39 (1.42) & 13.33 (6.35)\\
       & NRP-Int & 6.90 (1.29) & 7.85 (5.47) & 11.00 (5.45) & 23.45 (9.31) &  55.4 (10.84) & 23.75 (6.37)\\
       & RNP-Meta  & 2.10 (0.45) & 0.85 (1.04) & 16.75 (1.48) & 6.00 (2.90) & 7.37 (3.01) & 14.01 (6.37) \\
       & RNP-Pool & 8.00 (0.00) & 19.00 (0.00) & 12.55 (1.43) & 5.00 (0.73) & 25.60 (1.19) & 22.40 (6.35) \\
        \specialrule{0em}{3pt}{3pt}
       Error 3 
       & RNP-Int & 6.85 (1.27) & 5.50 (2.54) & 13.35 (2.25) & 1.90 (1.77) & 10.89 (3.02) & 15.08 (1.70) \\
      Cauchy(0,1)  & NRNP-Int & 6.95 (1.28) & 14.85 (2.21) & 4.05 (2.20) & 32.50 (3.80) & 245.89 (44.16) & 77.03 (11.67) \\
       & RP-Int   & 7.05 (0.69) & 3.85 (3.00) & 15.45 (2.31) & 1.75 (1.45) & 7.92 (3.22) & 13.94 (1.79) \\
       & NRP-Int & 7.00 (1.12) & 14.55 (2.59) & 3.90 (0.45) & 31.65 (4.33) & 269.4 (50.36) & 73.30 (11.17) \\
       & RNP-Meta  & 3.75 (1.25) & 4.30 (2.36) & 11.80 (2.19) & 15.75 (3.95) & 20.58 (5.42) & 17.00 (1.76) \\
       & RNP-Pool & 8.00 (0.00) & 19.00 (0.00) & 10.25 (1.59) & 5.05 (2.04) & 27.92 (2.18) & 21.91 (1.66)\\
 \bottomrule 
   \end{tabular}
   \label{tab:sce3-4}
\end{table}

\section{Application to TCGA datasets}\label{sec4}
As a cancer omics program initiated by the National Institute of Health (NIH), TCGA publishes high-quality profiling data on multiple cancer types. The rich data provide a valuable opportunity to uncover the underlying genetic factor of cancers, and more importantly, to identify the heterogeneity and commonality in the effects of genes on outcomes across different stages, subpopulations, or subtypes of the same type of cancer. In this section, we analyze TCGA data on 
glioblastoma multiforme (GBM) and lung adenocarcinoma (LUAD). The genetic and clinical data are downloaded from the cBioPortal website (http://www.cbioportal.org/) via the \emph{cgdrs} package. We use the processed level-3 gene expression data. Please refer to literature \cite[]{Dewey2011RSEM} for detailed information on the generation and processing of gene expression data. 

\subsection{GBM data} 
Glioblastoma multiforme (GBM) is the most common primary brain tumor of adults, as well as the most malignant. The response of interest is overall survival, which is right-censored. In the literature, it has been suggested that male patients have pathogenesis different from that of females \cite[]{Ostrom2018FemalesHT}, which is partially confirmed by our preliminary analysis (see Figure \ref{fig:scatter}). However, there is a lack of attention to the difference in some of the existing studies. As such, we divide the patients into two groups: male and female. After excluding subjects with missing in survival outcomes and gene expression, the male group has 158 samples with 133 deaths, and the female has 99 samples with 78 deaths. These two groups are expected to have both heterogeneity (as suggested by the literature) and commonality (as the two groups suffer from the same type of cancer). A total of 16, 270 RNAseq gene expressions are initially available for the analysis.  To improve the reliability of the analysis results, we reduce the gene set to the top 200 genes most correlated with the survival time by applying the {\color{black} nonparametric independence screening \cite[]{Fan2011} with normalized cubic B-spline as the basis function.} It is noted that this step, which excludes the most irrelevant genes, is a common practice for analyzing microarray data in cancer-related studies, as the number of cancer-related genes is not expected to be large. We rescale the gene expressions (covariates) such that their values have zero means and unit variances.  

\begin{figure}[t]
\centerline{\includegraphics[width=1\textwidth]{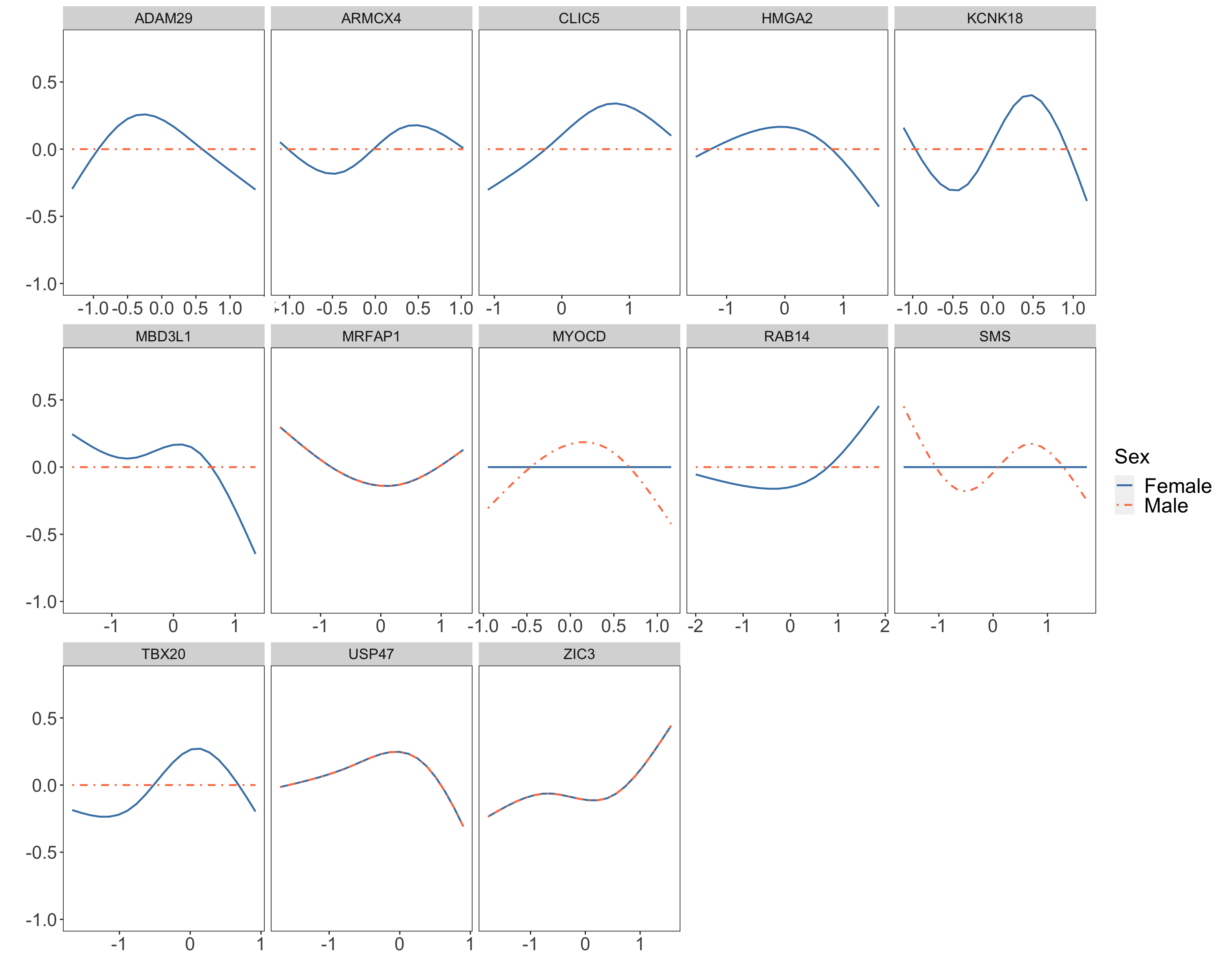}}
\caption{Analysis of GBM data using the proposed RNP-Int approach: estimated effects of 13 genes. Blue solid lines correspond to females, and orange dotted lines to males.\label{fig:gbm} }
\end{figure}

The nonlinear effect of some genes on the survival time has been observed in Figure \ref{fig:scatter}, which motivates the nonlinear modeling of gene expressions. We adopt the proposed RNP-Int method for this data under the AFT model, of which the tuning parameter $\lambda$ is selected by five-fold CV, in the same way as in simulation. Results are briefly presented in Figure \ref{fig:CV} (Appendix C). With RNP-Int, 13 genes are identified. Among them, three genes behave the same for both males and females. Figure \ref{fig:gbm} presents the estimated gene effects. It is observed that most of the estimated effects diverge from a straight line, suggesting that the nonparametric modeling is more appropriate to describe the gene effects on survival time in our data. 
A quick literature search suggests that the identified genes have important biological implications. For example, gene ZIC3 is identified for two subgroups. It is a positive regulator of host gene ARRB1 of miR-326, a GBM downregulated miRNA, which may partly lead to the downregulation of ARRB1 and miR-326 in GBM \cite[]{Nawaz2016PI3KP}. We also identify eight genes that are associated with the response in females but not males. Among them, gene ADAM29 has been found to be a GBM-specific gene in the de novo pathway of primary GBM, and it has been suggested as a viable target for the treatment of GBM \cite[]{Rahane2018ACT}. Another gene that we find to be associated with survival time in females only is gene HMG2, which is a transcriptional modulator that mediates motility and self-renewal in normal and cancer stem cells. It has been found that the level of HMGA2 expression increases in the majority of primary human GBM tumors compared to a normal brain, as well as in CD133+ GBM neurosphere cells compared to CD133- cells\cite[]{Kaur2016TheTM}. Further, gene HMGA2 is strongly associated with survival in GBM, suggesting it as a promising biomarker for predicting the survival time of GBM patients \cite[]{Liu2010PolymorphismsOL}. Genes that have significant effects in males but not females include MYOCD and SMS. Gene MYCOD encodes a nuclear protein, which is expressed in the heart, aorta, and smooth muscle cell-containing tissues. It may play a crucial role in cardiogenesis and differentiation of the smooth muscle cell lineage \cite[]{}. Gene SMS, a polyamine biosynthetic enzyme, has been reported to be over-expressed in colorectal cancer \cite[]{Guo2020Spermine}. 

\renewcommand\arraystretch{1.0}
 \begin{table}

\caption{Data analysis: number of overlapping genes. In each cell, male/female for GBM data, and stage I/II/III and IV for LUAD data}
 \begin{center}
    \setlength{\tabcolsep}{2.5pt}
    \renewcommand{\arraystretch}{1.0}
	\begin{tabular}{lccccccc}
    \hline
 & RNP-Int & NRNP-Int & RP-Int & NRP-Int & RNP-Meta & RNP-Pool \\ \hline
 \multicolumn{7}{c}{GBM}     \\
RNP-Int & 5/11 & 3/5 & 2/2 & 2/2 & 2/3 & 5/3 \\
NRNP-Int &  & 9/9 & 2/4 & 2/3 & 3/3 & 3/1 \\
RNP-Meta &  &  & 5/9 & 4/6 & 0/2&  2/0 \\
RNP-Pool  &  &  &  & 6/11 & 0/2 & 2/0 \\ 
RP-Int  & & & & &  8/10& 2/0 \\
NRP-Int  & & & & & & 5/5 \\\hline 
 \multicolumn{7}{c}{LUAD}\\
RNP-Int & 14/15/16 & 8/7/6 & 2/3/3 &  1/1/3&  3/3/7  & 14/15/15 \\
NRNP-Int &  & 18/13/14 & 1/1/1 & 1/0/0 & 3/3/5 & 8/7/6 \\
RP-Int  &  &  & 10/10/11& 3/2/4 &  1/3/1 & 3/3/3 \\
NRP-Int  & & & & 14/13/17 & 2/3/2 & 2/1/3 \\
RNP-Meta & & & & &  17/17/16 & 3/3/6 \\
RNP-Pool & & & & &  & 16/16/16 \\
\hline
 	\end{tabular}
\end{center}
\label{tab:compare1}
\end{table}

\renewcommand\arraystretch{1.0}
 \begin{table}

\caption{Data analysis: number of identified genes, heterogeneities, and commonalities.}
 \begin{center}
    \setlength{\tabcolsep}{5pt}
    \renewcommand{\arraystretch}{1.0}
	\begin{tabular}{lccc}
    \hline
 & Genes & Heterogeneities & Commonalities \\ \hline
 \multicolumn{4}{c}{GBM}     \\
RNP-Int & 13& 10 & 3  \\
NRNP-Int & 16  & 15 &  1   \\
RP-Int  &12 & 10  &  2 \\
NRP-Int  &  15 & 14 &  1 \\
RNP-Meta & 18 & 18 & 0  \\
RNP-Pool  & 5  &  0 & 5 \\ 
\hline 
 \multicolumn{4}{c}{LUAD}\\
RNP-Int & 17 & 5 & 12  \\
NRNP-Int & 18 & 6  & 12  \\
RP-Int  & 11 & 1 &  10  \\
NRP-Int &  17 & 4 &  13 \\
RNP-Meta &  40   & 40    &  0  \\
RNP-Pool  & 16 &  0  &  16  \\ 
\hline
 	\end{tabular}
\end{center}
\label{tab:compare2}
\end{table}

Data are also analyzed using alternative methods. The comparative results are provided in Table \ref{tab:compare1} and \ref{tab:compare2}. It is noted that different methods generate different results. NRNP-Int, RP-Int, and NRP-Int identify 16, 12, and 15 genes, respectively. RNP-Meta identifies 18 genes, and
all genes with nonzero effects are identified as behaving differently across two datasets. RNP-Pool identifies five genes. With its particular property, the effects of the identified genes are the same in the two datasets. Fewer overlaps between the proposed RNP-Int and RP-Int, as well as NRP-Int, in the set of identified genes in two datasets, suggest that genes have non-negligible nonlinear effects on survival times. Omitting these can significantly affect the identification results. 
Detailed estimation results using these alternative methods are provided in Figures \ref{fig:gbm_nrnpi}-\ref{fig:gbm_rnpp} (Appendix C).  

Based on real data, the true set of important genes is unknown and, thus, it is challenging to evaluate the identification and estimation accuracy directly. To verify the results, we now evaluate the prediction. Specifically, each dataset is randomly partitioned into a training set and a testing set, at a ratio of 4:1. 
The penalty parameters for the proposed RNP-Int and alternatives NRNP-Int, RP-Int, and NRP-Int are selected by five-fold CV within the training sample. Estimation (and tuning parameters) is conducted with the training set, and prediction is made with the testing set. With the survival data, we compute the log-rank statistics. Higher log-rank statistics indicate better prediction performance. {\color{black}The average values of log-rank statistics over 100 random replicates are 7.783 (RNP-Int), 4.479 (NRNP-Int), 6.559 (RP-Int), 4.736 (NRP-Int), 4.921 (RNP-Meta), and 5.239 (RNP-Pool), respectively.} The proposed approach has the highest log rank and thus has the best prediction performance.

\subsection{LUAD data} 
Lung adenocarcinoma (LUAD) is a form of non-small cell lung cancer, which is the most common type of lung cancer. In our analysis, we are interested in the regulation of FEV1 (forced expiratory volume in 1 second, prebroncholiator), a critical measure of lung function, by gene expressions. The left panel of Figure \ref{fig:scatter_LUAD} (Appendix C) shows the scatterplot of FEV1. It can be observed that there are indeed some outliers that are outside the three-standard-deviation range. The right panel of Figure \ref{fig:scatter_LUAD} (Appendix C) presents the plot of FEV1 against the expression of gene {\color{black} ARHGAP4}, as well as regression curves and bootstrap-based 95\% confidence intervals. It is easy to see that, for different stages of LUAD, the effects of genes on FEV1 are significantly different, and the effects are likely to be nonlinear. However, some of the existing analyses use samples of all tumor stages with insufficient attention to the potential heterogeneity across different stages and also neglect the possible nonlinear effects of gene expression on the response variable. After removing missing values from the FEV1, a total of 225 patients, with 139 in stage 1, 52 in stage II, and 34 in stages III and IV are included in the analysis. A total of 18, 325 RNAseq gene expression measurements are available. Using the same marginal screening process as described above, the number of gene expressions is reduced to 200. All gene expressions are rescaled to a range of 0-1. 

Under optimal tuning, the proposed RNP-Int method identifies 17 genes in at least one dataset. Among them, 12 genes behave the same across the three stages, and the rest behave differently. The estimated gene effects are shown in Figure \ref{fig:LUAD} (Appendix C). We again observe that most of the estimated effects may not have a simple linear approximation. A literature review suggests that the findings are biologically meaningful. For example, gene IL31 
is identified in all three stages. It has been known that IL31 has important effects on the pathogenesis of allergic diseases and tumors. One genetic variant of IL31, rs4758680, is found to be significantly associated with metastasis and poor survival status in LUAD \cite[]{Yang2018TheRO}. ADIPOQ is a protein-coding gene. The encoded protein circulates in the plasma and is involved with metabolic and hormonal processes. The TT genotype of rs2241766, a single-nucleotide polymorphism (SNP) in the gene ADIPOQ, is significantly associated with susceptibility to non-small cell lung cancer and is also related to the overall survival of non-small cell lung cancer patients \cite[]{Cui2011TheRO}. In addition to the genes associated with FEV1 in all three stages, we have also identified genes that have strong effects only in the earlier or later stages. Gene ARHGAP4 is found to have strong effects only in stages III and IV, while gene DCLK2 is found to be stage I specific. Genes that are associated with FEV1 in later stages are FRG2C and CHRM3. Gene FRG2C has been found in the non-ground-glass opacity LUAD, and it has been indicated to have significant biological functions related to the cell cycle and proliferation \cite[]{Kim2020GeneticFO}. Gene CHRM3 controls smooth muscle contraction and its stimulation causes secretion of the glandular tissue. The chemotherapy drug docetaxel resistant human NSCLC (A549) cell shows increased expression of CHRM3 \cite[]{Bland2020RepurposingAT}.  

We also apply the alternative methods described above to this data. The summary comparison results in Table \ref{tab:compare1} and \ref{tab:compare2} again suggest that different methods produce diverse results. Detailed estimation results using the alternatives are provided in \ref{fig:lu_nrnpi}-\ref{fig:lu_rnpp} (Appendix C). Prediction is evaluated as described above. With the continuous data, we compute the mean absolute error (MAE). A lower MAE indicates better prediction performance and {\color{black} the average MAE over 100 random replicates are 2.381 (RNP-Int), 2.867 (NRNP-Int), 2.547 (RP-Int), 2.960 (NRP-Int), 3.114 (RNP-Meta), and 2.406 (RNP-Pool), respectively.}The proposed approach is again observed to have better prediction performance.   

\section{Conclusions}\label{sec5}
In the current biomedical field, collected samples often span multiple subgroups. It is an important yet challenging problem to unveil the underlying heterogeneity and commonality across these related but different subgroups. In this paper, we propose a novel integrative analysis approach called RNP-Int for high-dimensional regression under multiple subgroups/datasets settings. RNP-Int advances from the existing literature by adopting a Cauchy loss function to accommodate possible data contamination and builds a nonparametric additive model to describe the effects of covariates in a more flexible manner. The proposed approach follows the sparse boosting framework, which has an appealing performance in high-dimensional data analysis. Significantly advanced from most of the existing sparse boosting approaches for multiple datasets, the proposed RNP-Int can identify heterogeneity and commonality across multiple datasets, as well as adopt a robust loss function and robust criteria. Simulation has demonstrated that RNP-Int outperforms five direct competitors in the identification of heterogeneity and commonality, variable selection, effect estimation, and prediction. In the analysis of two TCGA datasets, RNP-Int has led to sensible findings, which are significantly different from the alternatives. 

In this study, we have adopted the Cauchy loss function to accommodate possible data contamination. Multiple robust methods have been developed for high-dimensional data. However, no method is particularly better than others are. Recently, Barron \cite[]{Barron2019CVF} proposed a general robust loss function, which is a superset of many common robust loss functions, for example, Huber, least-squares, and Cauchy loss functions. By introducing robustness as a continuous parameter, this loss function can adaptively determine its specific form during the training process, which is particularly effective in real data modeling. It would be interesting to extend the current robust loss function with fixed form to this more general and adaptive robust loss function. Beyond the additive model considered in this study, 
there are many other nonparametric models. It will also be of interest to extend the proposed analysis to other more complex nonparametric models, for example, single- and multi-index models. Finally, it has been recognized that not only genes, but also gene-gene interactions play fundamental roles in understanding, modeling, and treating complex diseases. Interaction analysis has been conducted in genetic studies, but most of the existing attention has been on a single dataset. The integrative interaction approach with special attention to identifying heterogeneity/commonality across multiple datasets is still limited and expected to be very challenging. As considerable new developments will be needed, we postpone this interesting extension to future research. 


\section*{Acknowledgments}
This study was supported by the Fund for Building World-class Universities (disciplines) of Renmin University of China (18XNB004). The computer resources were provided by the Public Computing Cloud Platform of Renmin University of China.

\subsection*{Conflict of interest}

The authors declare no potential conflict of interests.


\newpage
\appendix

\section{Accelerated Failure Time model\label{sec:app1}}
The AFT model is an alternative to the commonly used Cox model in survival analysis because of its simple form, lucid interpretation, and low computational cost. In dataset $m$, for the $i$th subject, denote $T_i^m$ as the survival time and $C_i^m$ as the censoring time. In the AFT model setting, the nonparametric additive model (\ref{eq:model2}) is specified as 
\[\log(T_i^m)=\sum_{j=1}^p\phi(X^m_{ij})\beta_j^m+\epsilon_i^m,\]
where $\epsilon_i^m$ is the random error with an unknown distribution function. We assume that data have been properly normalized so that the intercept can be omitted. Under right censoring, we observe $\{(y_i^m,\delta_i^m, X_{ij}^m), i=1,\ldots,n^m, m=1,\ldots,M\}$ where $y_i^m=\log(\min(T_i^m,C_i^m))$, $\delta_i^m=1(T_i^m\leq C_i^m)$, and $X_{ij}^m$ are the covariates associated with $y_i^m$. Assume that $(y_i^m,\delta_i^m, X_{ij}^m)$'s have been ordered according to $y_i^m$'s in an ascending order. 

We adopt the Kaplan-Meier weights to accommodate censoring. 
The Kaplan-Meier weights $\omega_i^m$ can be computed as 
\begin{eqnarray*}
\omega_1^m & = &\frac{\delta_{\color{black}1}^m}{n^m}, \\
\omega_i^m & = & \frac{\delta_i^m}{n^m-i+1}\prod_{k=1}^{i-1} \left(\frac{n^m-k}{n^m-k+1}\right)^{\delta_k^m}, i=2,\ldots,n^m. 
\end{eqnarray*}
To accommodate contamination in survival time, we adopt the weighted Cauchy loss function: 
\[\rho_c(\beta^m)=\sum_{i=1}^{n^m}\omega_i^m \log\left[1+(y_i^m-\sum_{j=1}^p\phi(X^m_{ij})\beta_j^m)^2/c^2\right], m=1,\ldots,M.\]

\newpage 

\section{Alternative approaches RNP-Meta and RNP-Pool used in numerical study}
RNP-Meta and RNP-Pool are two non-integrative approaches. RNP-Meta conducts sparse boosting under robust nonparametric modeling on each dataset separately, and then results are combined across datasets; RNP-Pool combines all datasets directly and applies the sparse boosting under robust nonparametric modeling. The algorithms of RNP-Meta and RNP-Pool are summarized in Algorithm \ref{algA1} and \ref{algA2}. 

\begin{algorithm}
\caption{RNP-Meta}\label{algA1}
\begin{algorithmic}[1]
\State {\bf Initialization}. Set $\bm\beta^{(1)}=\bm 0$. 
  \For{$m=1,\ldots,M$}
  \For{$t=1,\ldots,T$} 
  \State compute  
\[(\hat{s^m}, \hat{\gamma^m}) = \text{argmin}_{1\leq s \leq p, \gamma \in \mathbb{R}^K} \rho_c\left(\beta^{m(t)}+\sum_{k=1}^K \gamma_k 1_{(s-1)K+k}\right)+\frac{\log n^m}{n^m}\sum_{j=1}^p \chi\left(\beta_j^{m(t)}+\delta_{js}\gamma\neq 0\right).\]
\State Update $\beta^{m(t+1)}=\beta^{m(t)}+v\sum_{k=1}^K\hat{\gamma^m_k}1_{(\hat{s^m}-1)K+k}$, where $v$ is the step size as in Algorithm 1. 
\State Compute $F^{m(t)}=\rho_c(\beta^{m(t+1)})+\frac{\log n^m}{n^m}\sum_{j=1}^p\chi\left(\beta_j^{m(t+1)}\neq 0\right).$
  \EndFor
 \State Select the optimal stopping criterion by $\hat{t^m}=\text{argmin}_{1\leq t \leq T}F^{m(t)}$. 
 \State $\beta^{m(\hat{t^m})}$ is the final estimator of $\beta^m$. The strong learner for dataset $m$ is $L_i^m=g\left(\sum_{j=1}^p\phi(X_{ij}^m)\beta_j^{m(\hat{t^m})}\right)$ with $i=1,\ldots, n^m$. 
 \EndFor
\end{algorithmic}
\end{algorithm}

\begin{algorithm}
\caption{RNP-Pool}\label{algA2}
\begin{algorithmic}[1]
\State {\bf Initialization}. Combine $M$ datasets into one, and denote $Y=[(Y^1)^\top,\ldots,(Y^m)^\top]^\top\in\mathbb{R}^N$ as the combined responses, $\bm X=[(\bm X^1)^\top,\ldots, (\bm X^M)^\top]^\top \in \mathbb{R}^{N\times p}$ as the combined measurement matrix, and $\beta\in \mathbb{R}^{Kp}$ as the unknown coefficient vector. Set $\beta^{(1)}=0$. 
  \For{$t=1,\ldots,T$} 
  \State compute  
\[(\hat s, \hat \gamma) = \text{argmin}_{1\leq s \leq p, \gamma \in \mathbb{R}^K} \rho_c\left(\beta^{(t)}+\sum_{k=1}^K \gamma_k 1_{(s-1)K+k}\right)+\frac{\log N}{N}\sum_{j=1}^p \chi\left(\beta_j^{(t)}+\delta_{js}\gamma \neq 0\right).\]
\State Update $\beta^{(t+1)}=\beta^{(t)}+v\sum_{k=1}^K\hat{\gamma_k}1_{(s-1)K+k}$, where $v$ is the step size as shown in Algorithm 1. 
\State Compute $F^{(t)}=\rho_c(\beta^{(t+1)})+\frac{\log N}{N}\sum_{j=1}^p\chi\left(\beta_j^{(t+1)}\neq 0\right).$
  \EndFor
 \State Select the optimal stopping criterion by $\hat{t}=\text{argmin}_{1\leq t \leq T}F^{(t)}$. 
 \State $\beta^{(\hat{t})}$ is the final estimator of $\beta$. The strong learner for the combined dataset is $L_i=g\left(\sum_{j=1}^p\phi(X_{ij})\beta_j^{(\hat{t})}\right)$ with $i=1,\ldots, N$.  
\end{algorithmic}
\end{algorithm}

\clearpage 
\section{More figures and tables}


\begin{figure}[t]
\centerline{\includegraphics[width=1\textwidth]{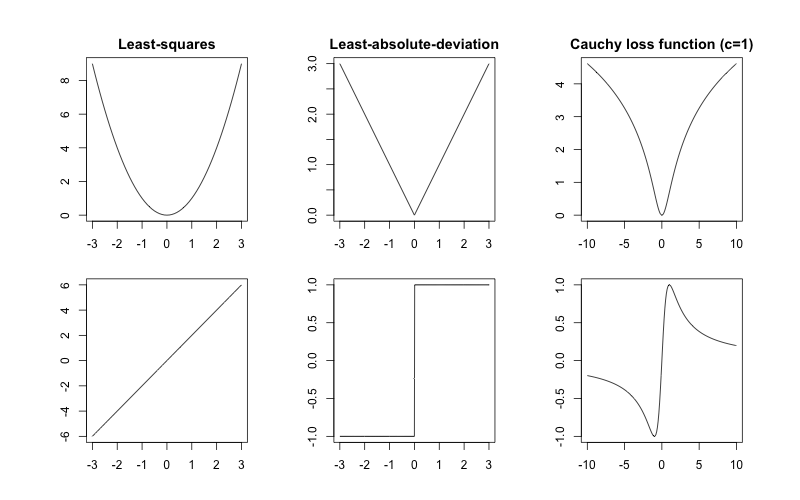}}
\caption{Illustration of different loss function. First row: loss functions $\rho(x)$. Second row: influence functions $\psi(x)$.\label{fig:CLF}}
\end{figure}

\begin{figure}[t]
\centerline{\includegraphics[width=1\textwidth]{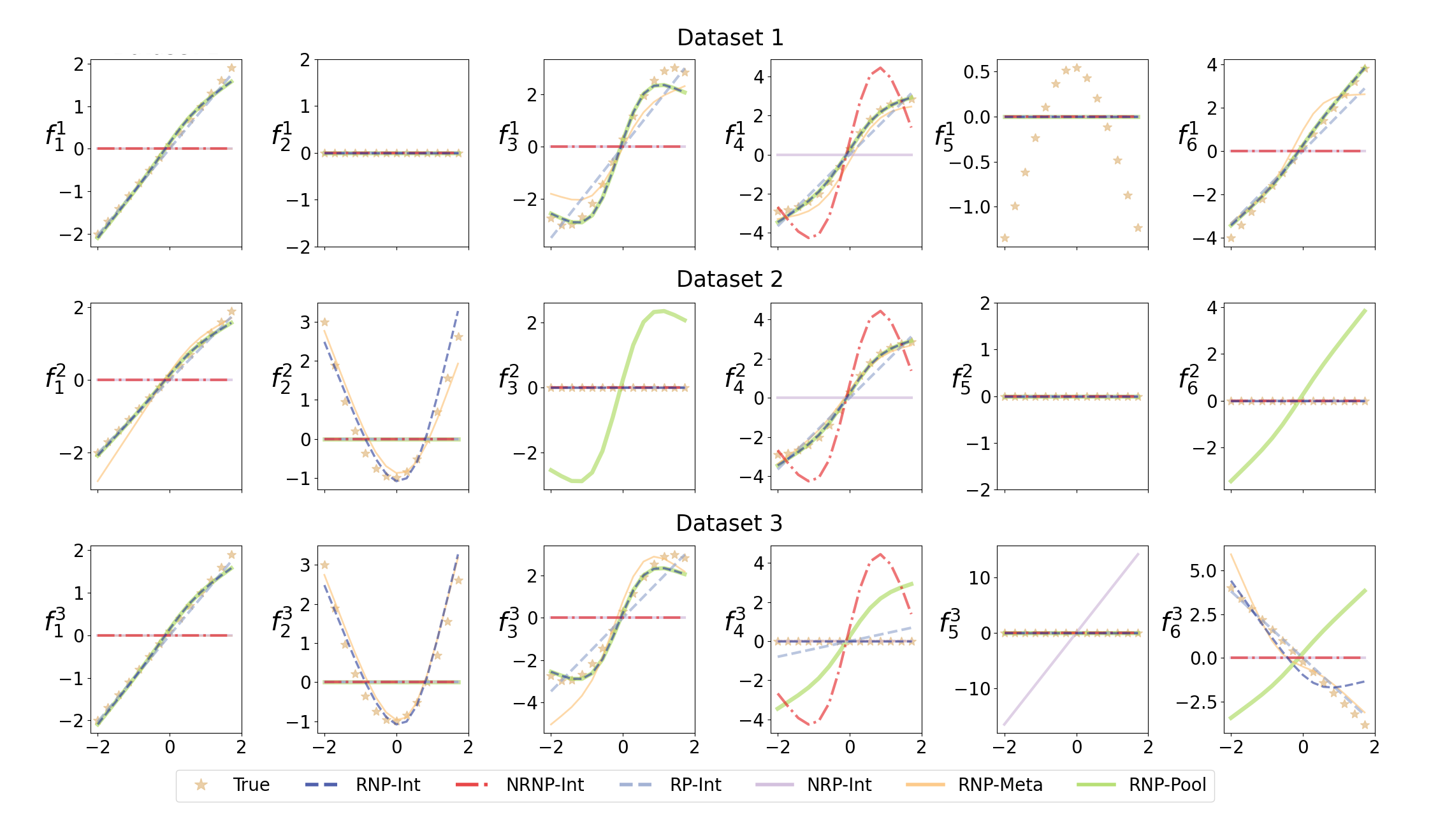}}
\caption{Simulation under Scenario 1 and error 3: estimation of the effects of the first six covariates. Three rows correspond to three datasets. \label{fig:sim}}
\end{figure}

\begin{figure}[t]
\centerline{\includegraphics[width=1\textwidth]{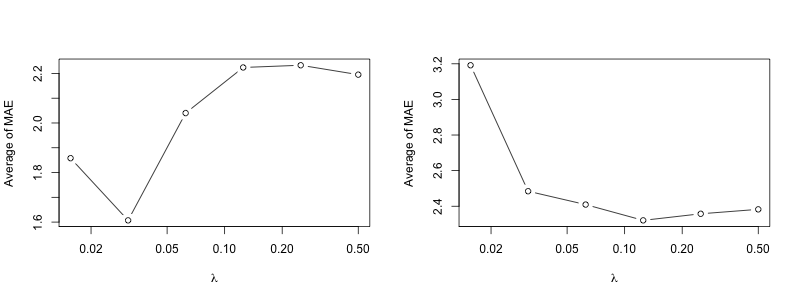}}
\caption{Data analysis: Average of MAE in five-fold CV as a function of $\lambda$. Left: GBM. Right: LUAD.\label{fig:CV}}
\end{figure}

\begin{figure}[t]
\centerline{\includegraphics[width=1\textwidth]{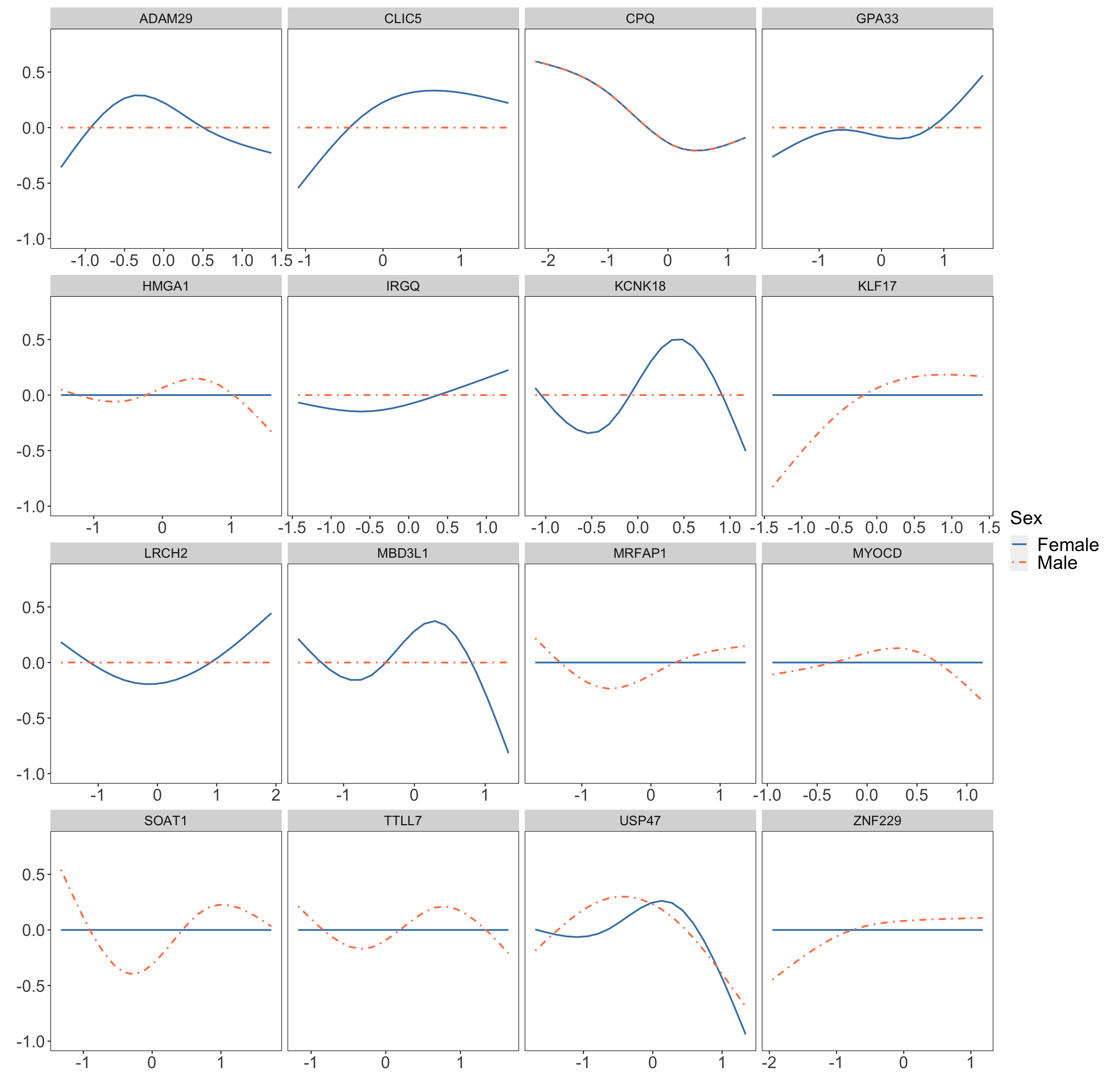}}
\caption{Analysis of GBM data using the alternative NRNP-Int method: estimated effects of 16 genes. Blue solid lines correspond to females, and orange dotted lines to males. \label{fig:gbm_nrnpi}}
\end{figure}

\begin{figure}[t]
\centerline{\includegraphics[width=1\textwidth]{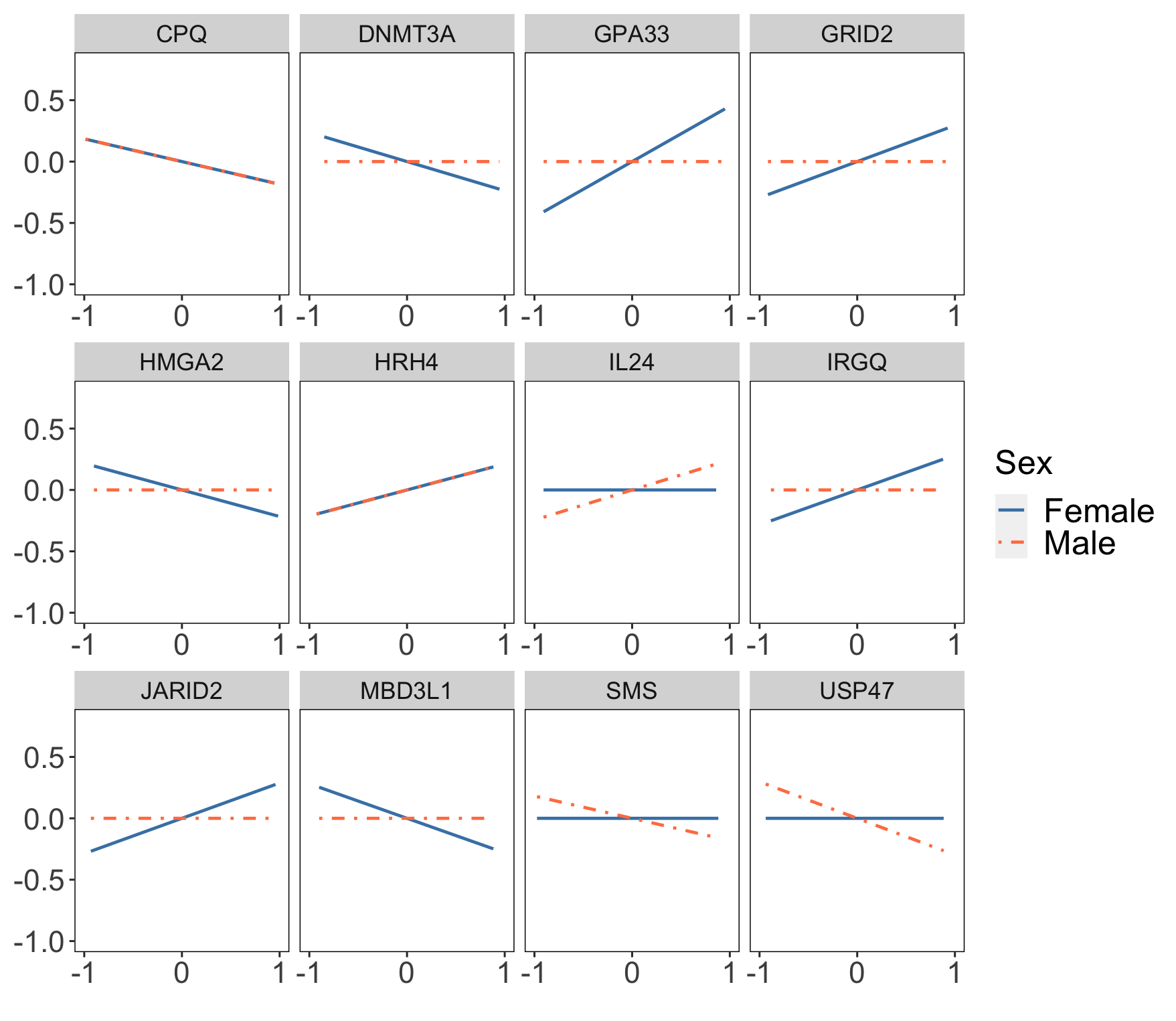}}
\caption{Analysis of GBM data using the alternative RP-Int method: estimated effects of 12 genes. Blue solid lines correspond to females, and orange dotted lines to males.\label{fig:gbm_rpi}}
\end{figure}

\begin{figure}[t]
\centerline{\includegraphics[width=1\textwidth]{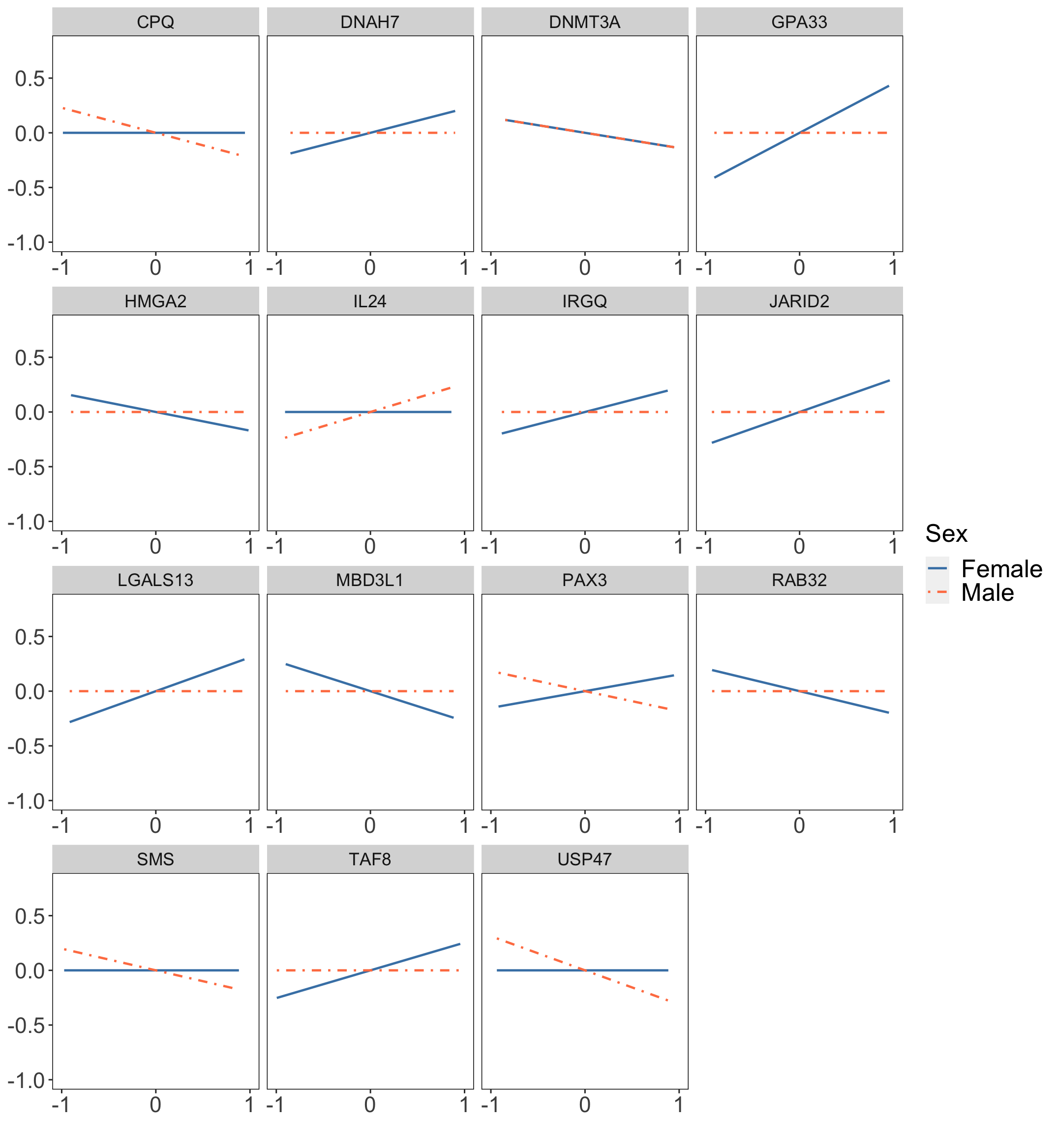}}
\caption{Analysis of GBM data using the alternative NRP-Int method: estimated effects of 15 genes. Blue solid lines correspond to females, and orange dotted lines to males.\label{fig:gbm_nrpi}}
\end{figure}

\begin{figure}[t]
\centerline{\includegraphics[width=1\textwidth]{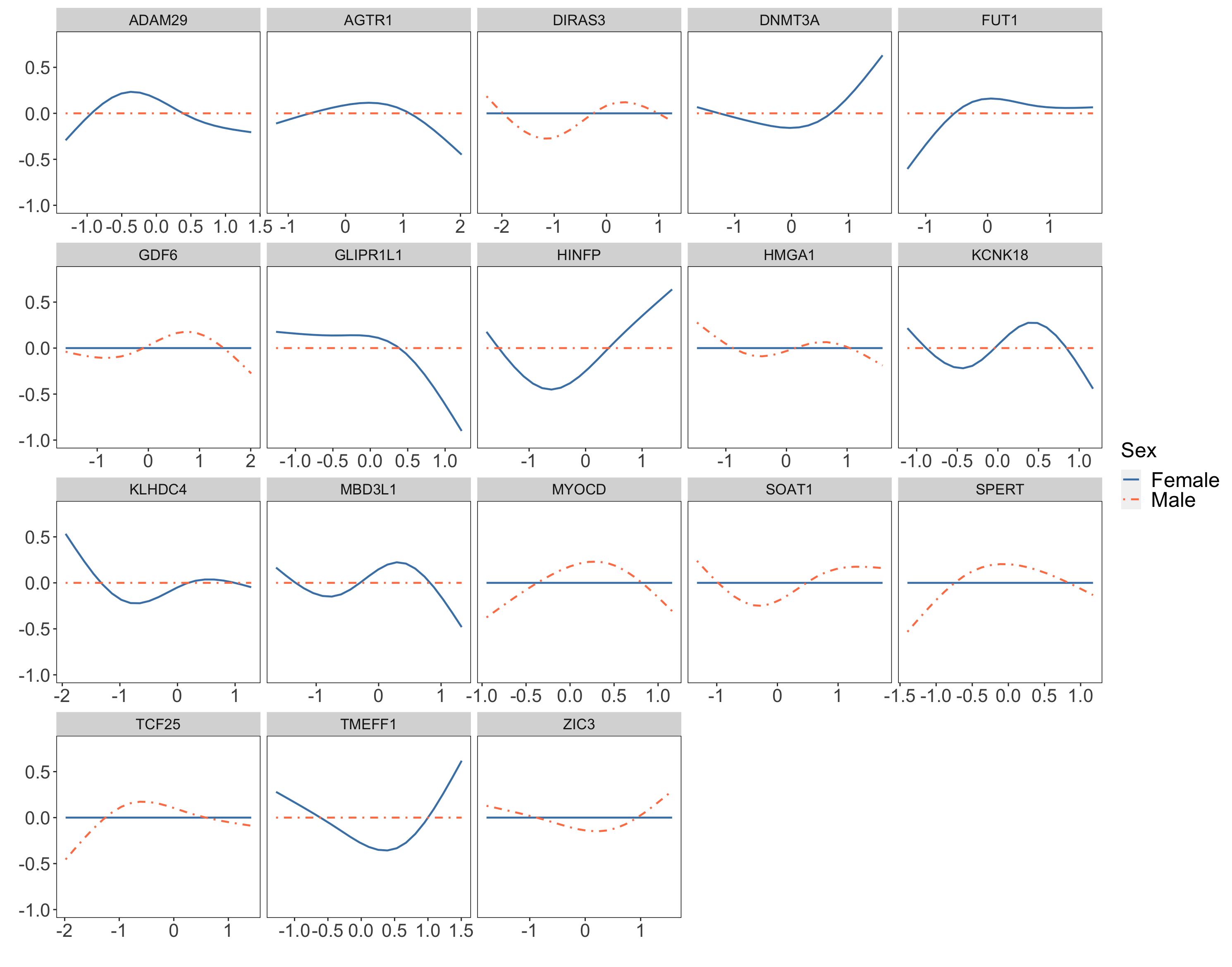}}
\caption{Analysis of GBM data using the alternative RNP-Meta method: estimated effects of 18 genes. Blue solid lines correspond to females, and orange dotted lines to males.\label{fig:gbm_rnpm}}
\end{figure}

\begin{figure}[t]
\centerline{\includegraphics[width=1\textwidth]{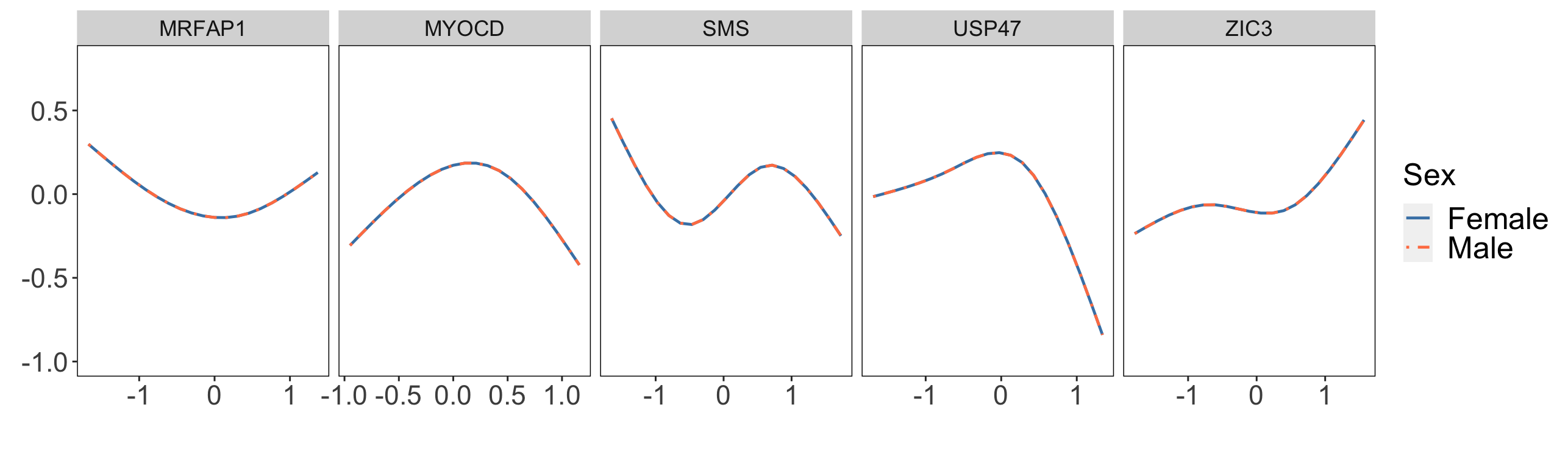}}
\caption{Analysis of GBM data using the alternative RNP-Pool method: estimated effects of 5 genes. Blue solid lines correspond to females, and orange dotted lines to males.\label{fig:gbm_rnpp}}
\end{figure}

\begin{figure}
\centerline{\includegraphics[width=0.9\textwidth]{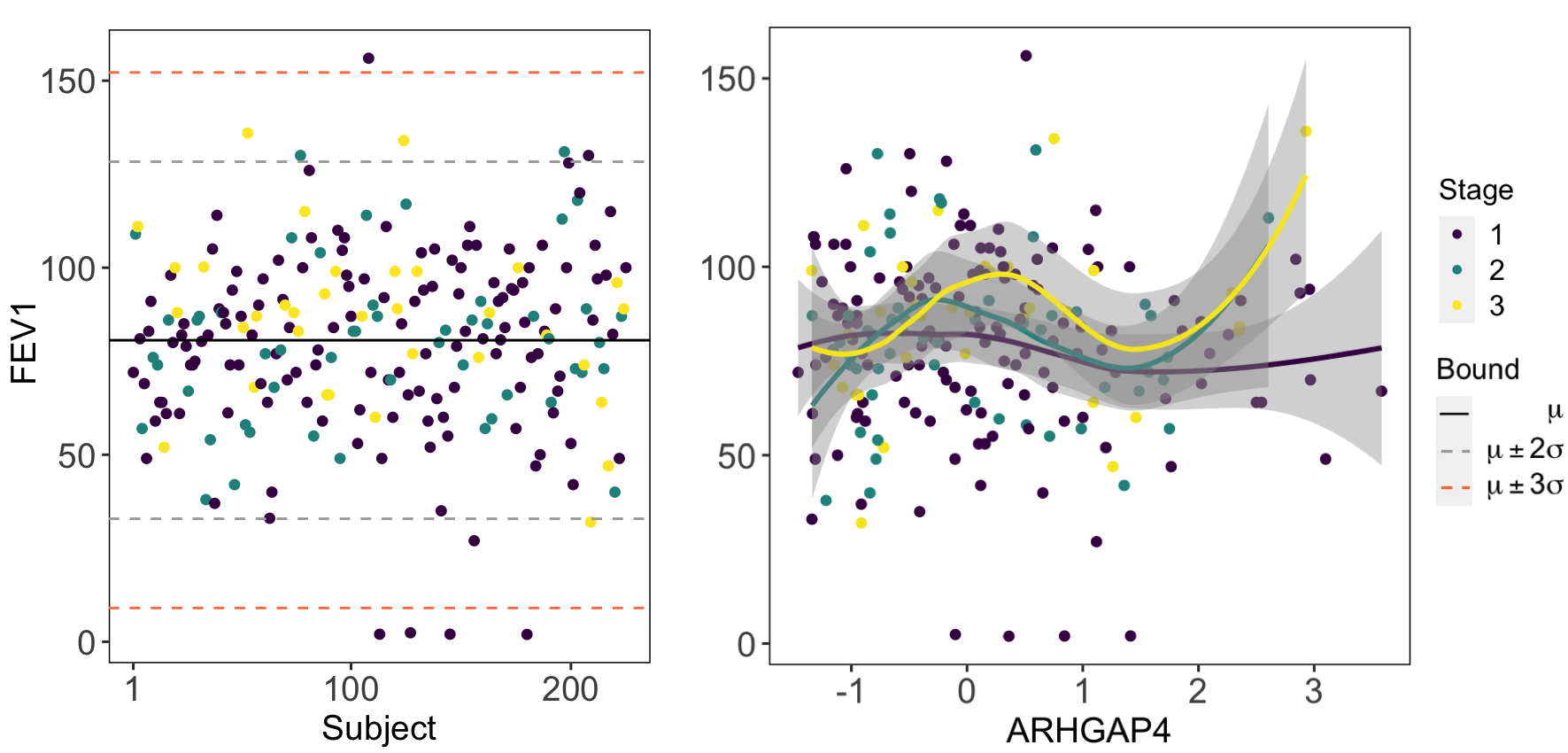}}
\caption{Analysis of TCGA LUAD data. Left: scatter plot of FEV1. Right: FEV1 against {\color{black}ARHGAP4} gene expression, and regression curve with boostrap-based $95\%$ confidence intervals (shaded area). \label{fig:scatter_LUAD}}
\end{figure}

\begin{figure}[t]
\centerline{\includegraphics[width=1\textwidth]{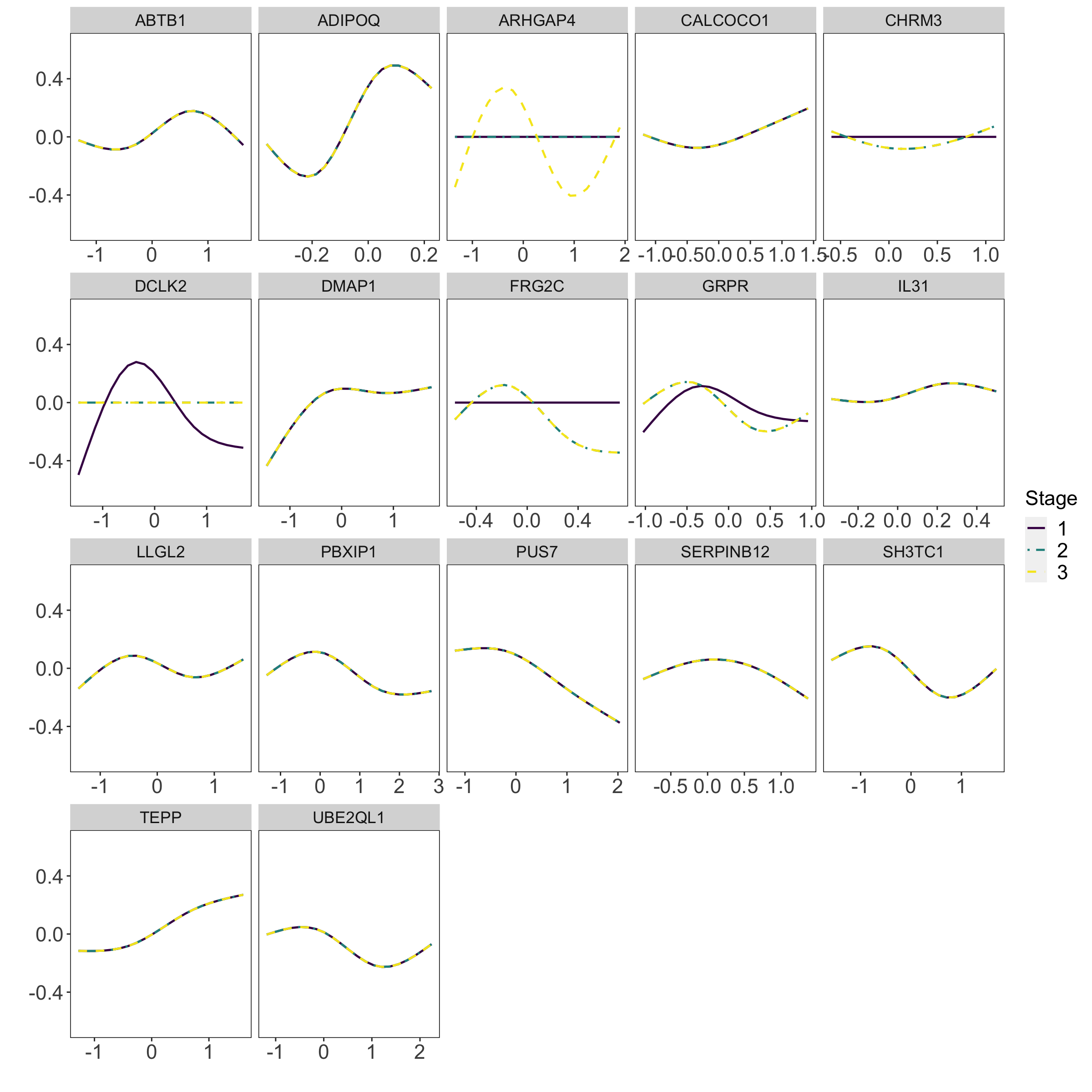}}
\caption{Analysis of LUAD data using the proposed RNP-Int method: estimated effects of 17 genes. Purple lines correspond to stage I, green lines to stage II, and yellow lines to stage III and IV. \label{fig:LUAD}}
\end{figure}

\begin{figure}[t]
\centerline{\includegraphics[width=1\textwidth]{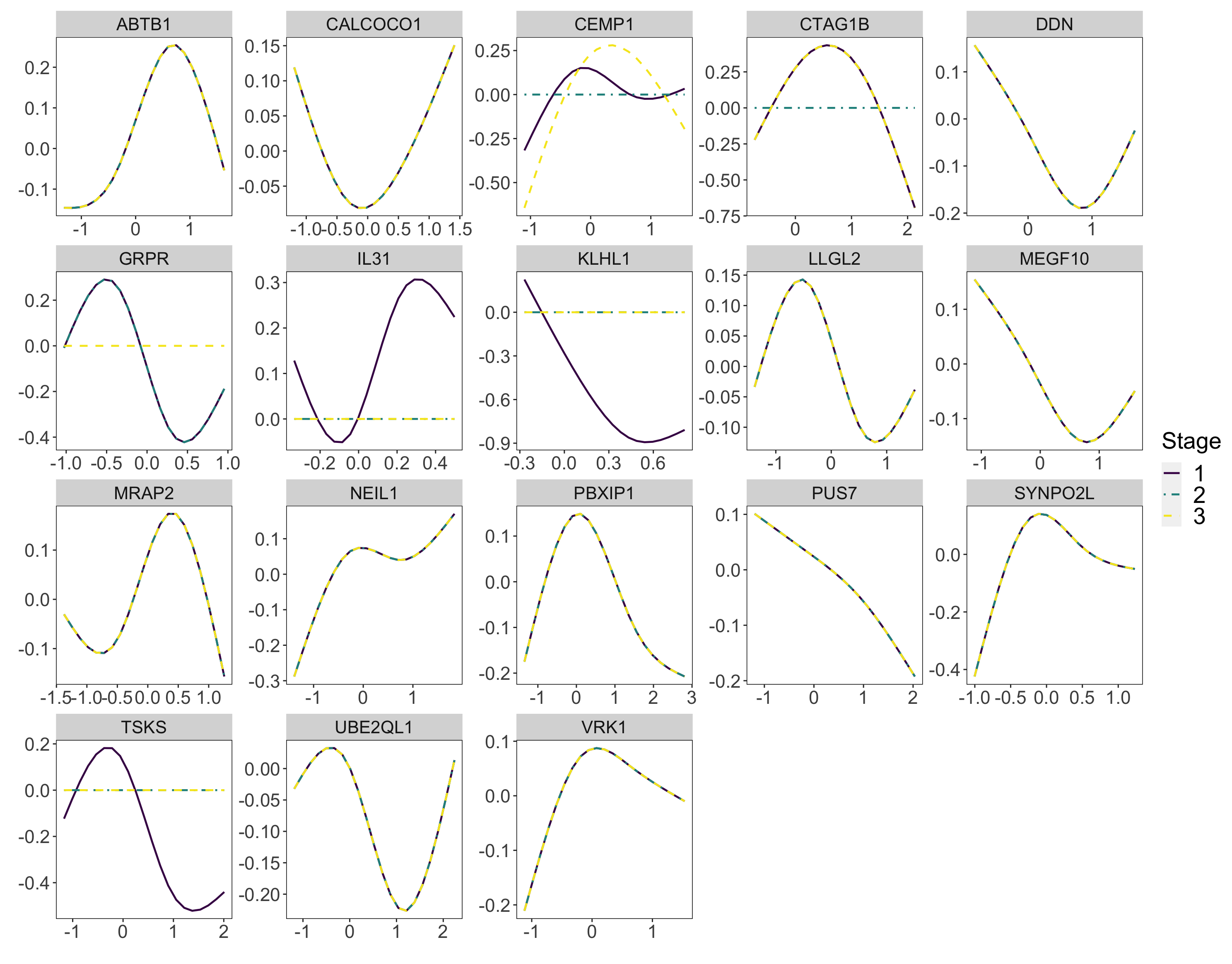}}
\caption{Analysis of LUAD data using the proposed NRNP-Int method: estimated effects of 18 genes. Purple lines correspond to stage I, green lines to stage II, and yellow lines to stage III and IV. \label{fig:lu_nrnpi}}
\end{figure}

\begin{figure}[t]
\centerline{\includegraphics[width=1\textwidth]{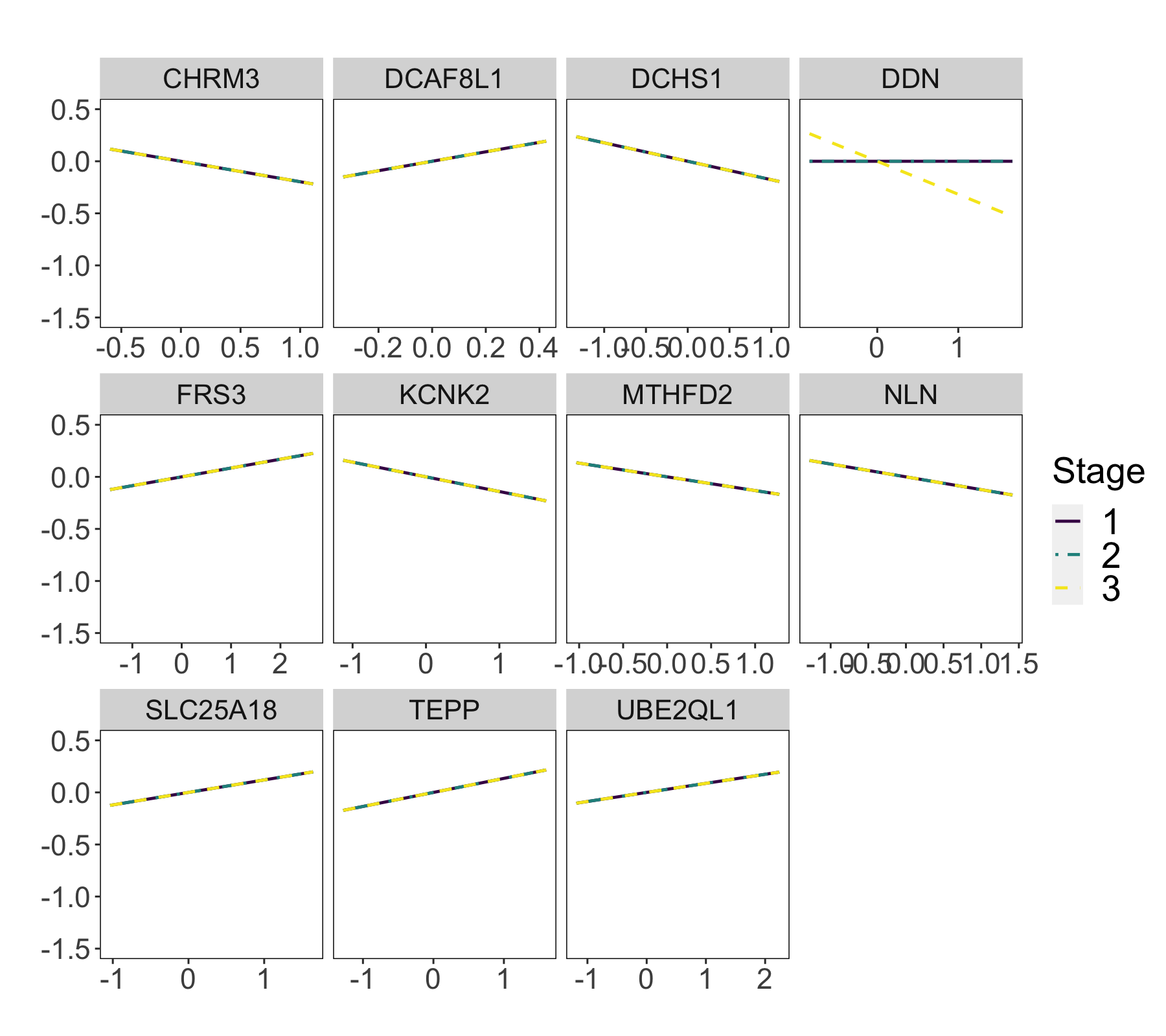}}
\caption{Analysis of LUAD data using the proposed RP-Int method: estimated effects of 11 genes. Purple lines correspond to stage I, green lines to stage II, and yellow lines to stage III and IV. \label{fig:lu_rpi}}
\end{figure}

\begin{figure}[t]
\centerline{\includegraphics[width=1\textwidth]{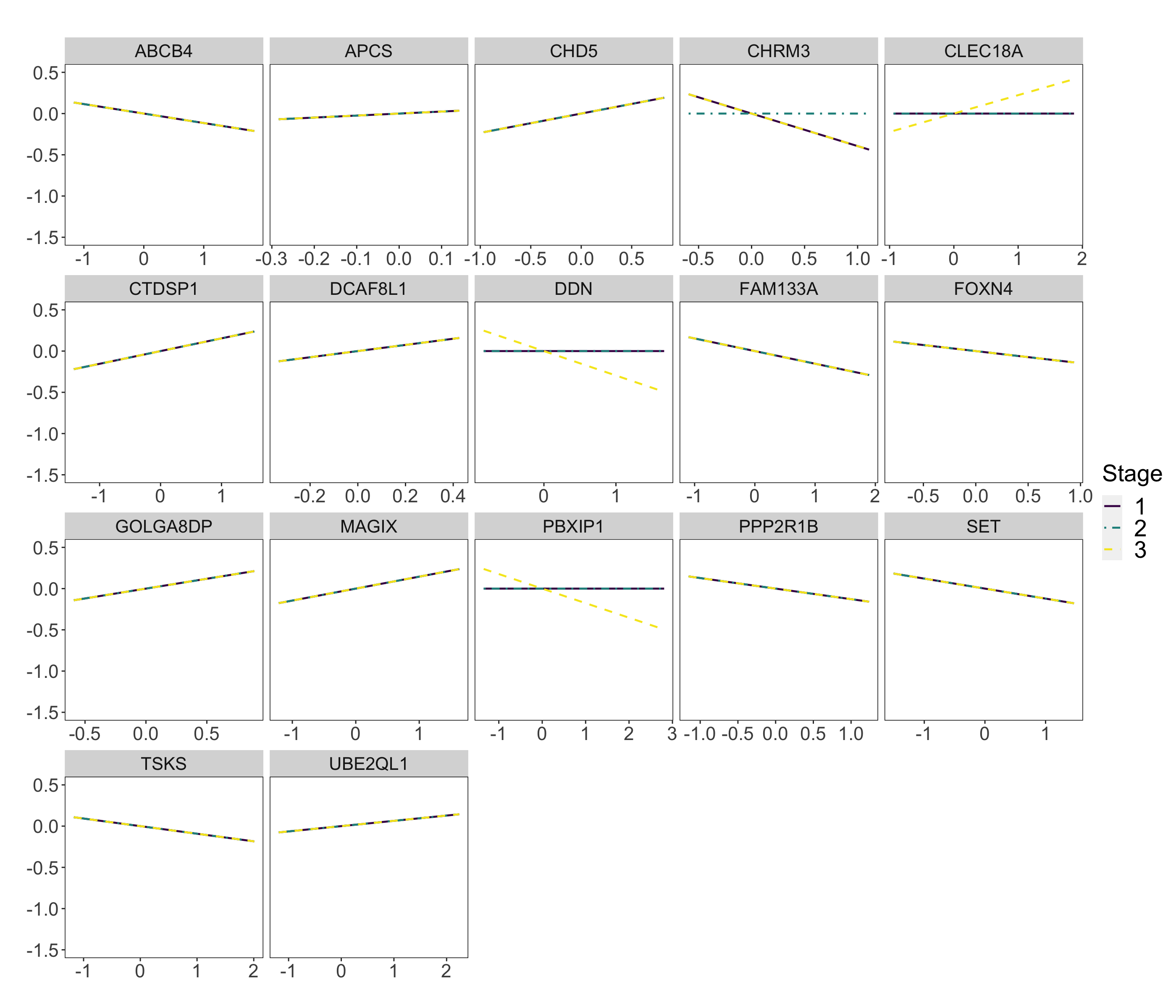}}
\caption{Analysis of LUAD data using the proposed NRP-Int method: estimated effects of 17 genes. Purple lines correspond to stage I, green lines to stage II, and yellow lines to stage III and IV. \label{fig:lu_nrpi}}
\end{figure}

\begin{figure}[t]
\centerline{\includegraphics[width=1\textwidth]{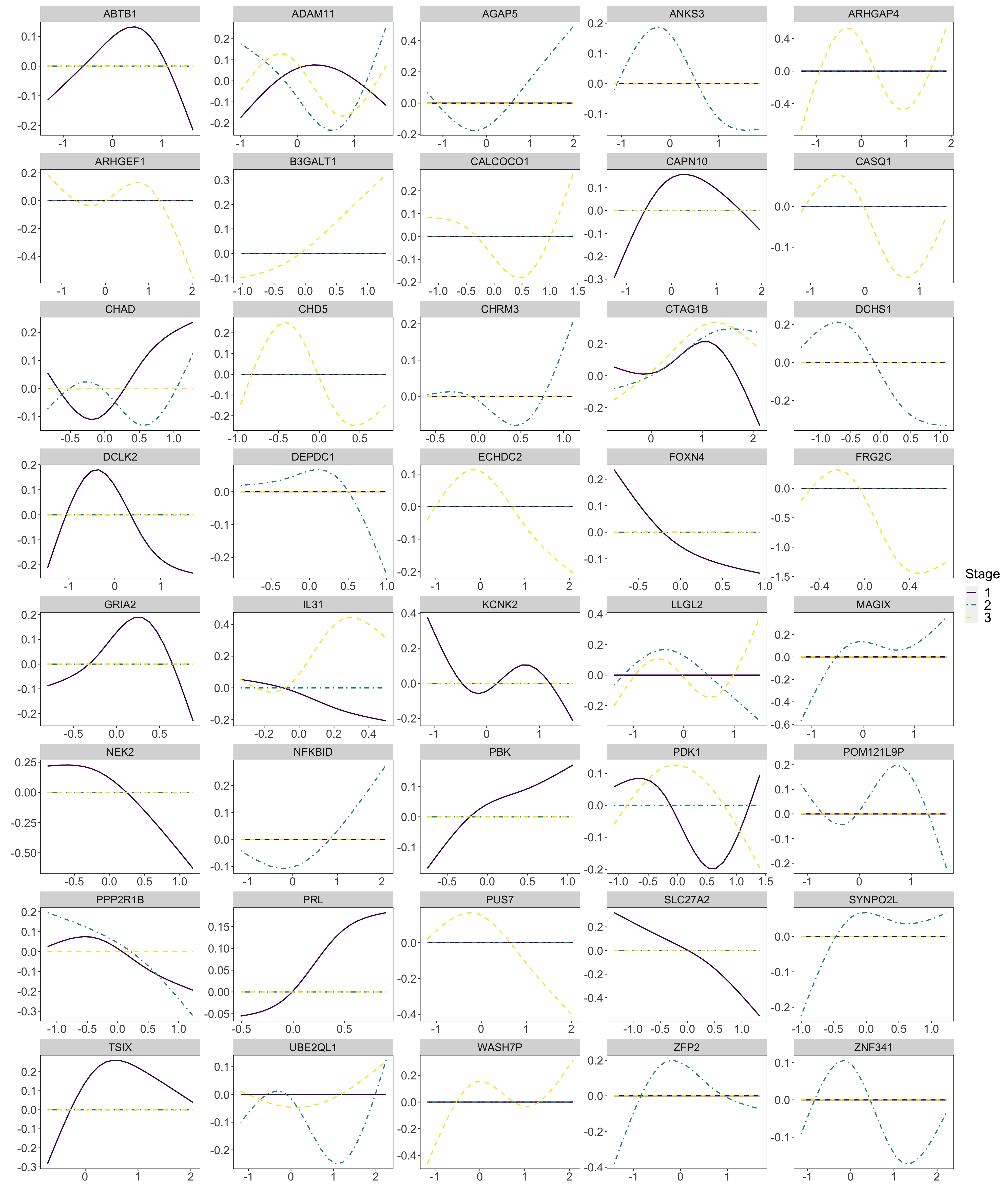}}
\caption{Analysis of LUAD data using the proposed RNP-Meta method: estimated effects of 40 genes. Purple lines correspond to stage I, green lines to stage II, and yellow lines to stage III and IV. \label{fig:lu_rnpm}}
\end{figure}

\begin{figure}[t]
\centerline{\includegraphics[width=1\textwidth]{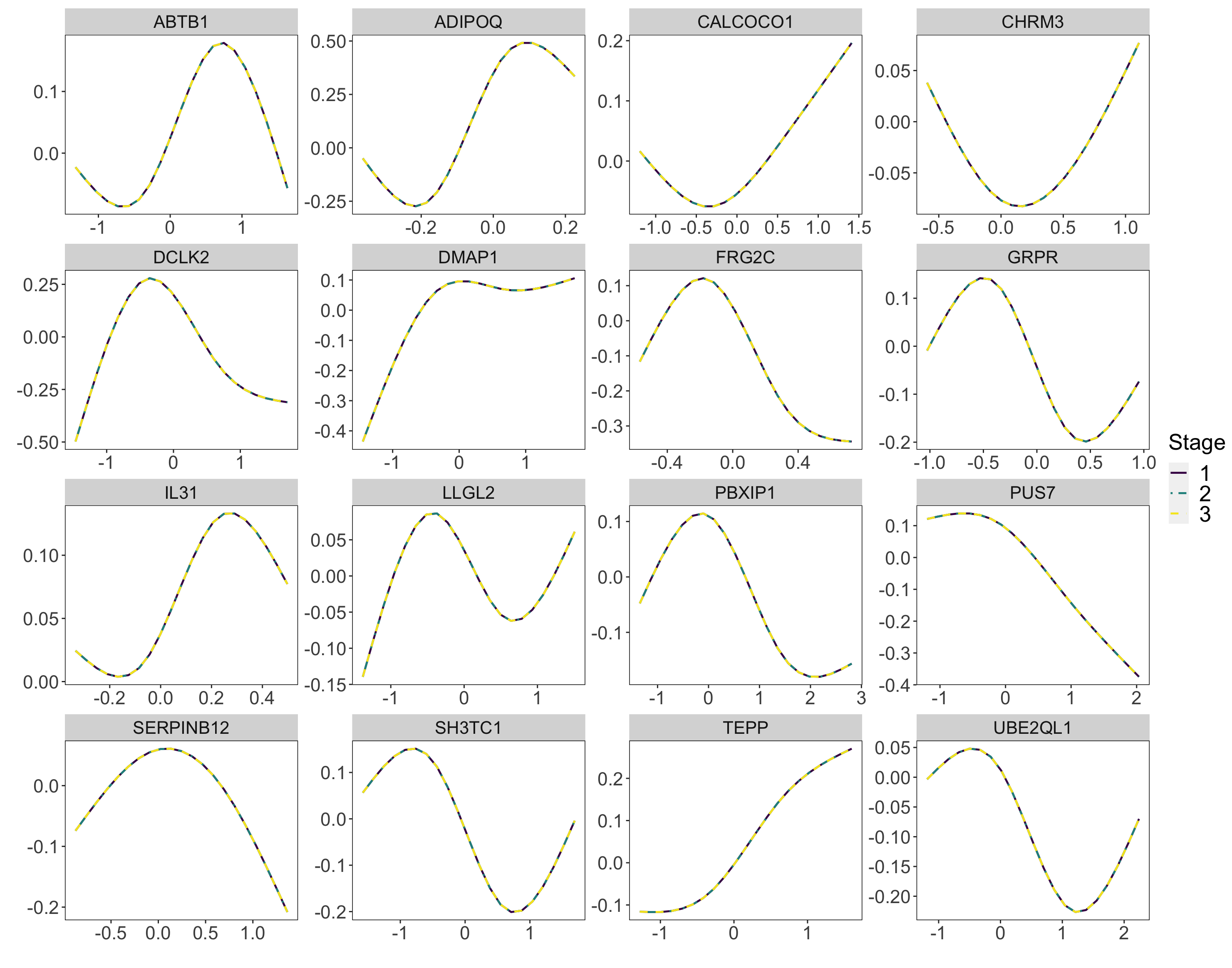}}
\caption{Analysis of LUAD data using the proposed RNP-Pool method: estimated effects of 16 genes. Purple lines correspond to stage I, green lines to stage II, and yellow lines to stage III and IV. \label{fig:lu_rnpp}}
\end{figure}

\begin{table}[htbp]
   \centering
   \caption{
   Simulation Scenarios 1-2 for data with a right-censored survival outcome. In each cell, mean (standard deviation).} 
   \begin{tabular}{llcccccc} 
      \toprule
      & & \multicolumn{2}{c}{\bf Commonality identification} & \multicolumn{2}{c}{\bf Variable selection}  \\
      \cmidrule(r){3-4} \cmidrule(l){5-6} 
       & \bf Methods & \bf TP-ind &\bf FP-ind & \bf  TP-var  &\bf FP-var & \bf RMISE & \bf {\color{black}Cstat} \\
      \midrule
       & & & & \bf{ Scenario 1} \\
       Error 1 & RNP-Int & 6.10 (1.17) & 1.75 (0.79)& 11.10 (0.45) & 0.30 (0.47)  & 2.58 (0.81) & 0.90 (0.01)\\
       N(0,1) 
        & NRNP-Int & 6.30 (0.87) & 0.50 (0.89)  & 11.80 (0.41)  & 1.65 (1.18) & 2.57 (0.42) & 0.90 (0.01) \\
       & RP-Int & 6.30 (1.03) & 4.30 (0.98) & 8.80 (0.62) & 0.40 (0.60) & 4.38 (1.17) & 0.85 (0.03) \\
       & NRP-Int  & 5.45 (1.40) & 3.45 (0.69) & 9.30 (0.57) & 13.10 (3.16) & 6.21 (1.20) & 0.82 (0.03)  \\ 
       & RNP-Meta  & 1.00 (0.32) & 0.85 (1.27) &11.20 (1.11) & 6.20 (2.35) & 5.62 (2.42) & 0.87 (0.04)  \\
       & RNP-Pool &  7.00 (0.00)& 11.00 (0.00)& 9.25 (0.55)   & 3.35 (0.75) & 10.80 (0.81) & 0.71 (0.03) \\
       \specialrule{0em}{3pt}{3pt}
       Error 2 & RNP-Int & 6.45 (0.95) & 2.05 (0.83) & 11.00 (0.32) & 0.5 (0.69) & 2.62 (0.98) & 0.86 (0.03)  \\
       70\%$N(0,1)$ & NRNP-Int & 5.45 (1.23) & 5.50 (3.36) & 6.85 (4.08) & 18.75 (5.16) & 48.37 (9.08) & 0.66 (0.11)\\
       +30\%Cauchy(0,1) & RP-Int & 6.05 (0.89) & 4.05 (0.39) & 8.90 (0.45) & 0.50 (0.76) & 4.23 (0.79) & 0.82 (0.03) \\
       & NRP-Int  & 5.70 (1.22) & 6.80 (3.00) & 5.30 (3.85) &20.85 (3.96) & 56.16 (10.52) & 0.63 (0.11) \\  
       & RNP-Meta  & 1.00 (0.32) & 1.95 (1.40) & 10.25 (1.25) & 9.50 (2.67) & 8.58 (3.37) & 0.81 (0.05)  \\
       & RNP-Pool & 7.00 (0.00) & 11.00 (0.00) & 9.15 (0.49) & 3.30 (0.80) & 10.83 (6.01) & 0.69 (0.03)  \\
       \specialrule{0em}{3pt}{3pt}
       Error 3 & RNP-Int & 6.00 (1.17)  & 3.30 (2.20) & 9.70 (2.45) & 2.10 (2.13) & 6.42 (4.79) & 0.76 (0.05) \\
       Cauchy(0,1) 
       & NRNP-Int & 6.45 (1.00) & 9.30 (2.03) & 1.70 (2.43) & 22.15 (3.00) & 237.62 (45.4) & 0.49 (0.07) \\
       & RP-Int   & 6.30 (0.80) & 4.55 (1.05) & 8.35 (1.14) & 0.85 (1.04) & 5.39 (1.76) & 0.75 (0.03) \\
       & NRP-Int & 5.95 (1.47) & 9.85 (1.46) & 1.00 (1.56) & 22.65 (3.01) & 202.38 (32.59) & 0.48 (0.08) \\
       & RNP-Meta  & 1.85 (0.88) & 4.20 (1.58) & 7.15 (1.63) & 16.50 (3.41) & 16.82 (4.37) & 0.69 (0.05)  \\
       & RNP-Pool & 7.00 (0.00) & 11.00 (0.00) & 8.25 (2.34) & 4.80 (2.80) & 14.20 (1.00)  & 0.64 (0.04) \\
       \midrule
        & & &  & \bf{Scenario 2} \\
        Error 1 & RNP-Int & 17.90 (0.45) & 0.00 (0.00) & 17.85 (0.67) & 0.00 (0.00) & 2.41 (0.62) & 0.92 (0.01)   \\
       N(0,1)  &NRNP-Int & 17.70 (0.73)  & 0.00 (0.00) & 18.00 (0.00) & 1.50 (1.10) & 2.78 (0.60)  & 0.92 (0.01)   \\ 
       & RP-Int   &  17.70 (0.73) & 0.00 (0.00) & 12.05 (0.39) & 0.05 (0.39) & 6.01 (0.39) & 0.87 (0.02)\\
       & NRP-Int &  15.25 (1.48) & 0.00 (0.00) & 12.35 (1.04) & 17.65 (3.51) & 13.17 (1.43) & 0.84 (0.02)\\
       & RNP-Meta  & 1.75 (2.00) & 0.00 (0.00) & 14.35 (2.28) & 9.10 (3.85) & 11.41 (4.19) & 0.85 (0.04)   \\
       & RNP-Pool & 18.00 (0.00) & 0.00 (0.00) & 17.85 (0.67) & 0.00 (0.00) & 2.38 (0.60) & 0.91 (0.01)   \\
        \specialrule{0em}{3pt}{3pt}
       Error 2   
       & RNP-Int & 17.90 (0.45) & 0.00 (0.00)  & 17.85 (0.67) & 0.15 (0.67)  & 2.62 (0.82) & 0.89 (0.02) \\
      70\%$N(0,1)$ 
       & NRNP-Int &  15.50 (2.33) & 0.00 (0.00) & 10.05 (5.41) & 18.80 (3.93) & 52.23 (9.52) & 0.71 (0.11) \\
      +30\% Cauchy(0,1)  & RP-Int  &  17.20 (1.20) & 0.00 (0.00) & 12.20 (0.77) & 0.30 (0.73) & 6.21 (0.57) & 0.83 (0.02)\\
       & NRP-Int & 15.80 (2.33) & 0.00 (0.00) & 7.70 (4.47) & 20.25 (3.37) & 59.8 (11.32) & 0.67 (0.12) \\
       & RNP-Meta  & 3.35 (0.18) & 0.00 (0.00) & 11.30 (2.62) & 14.60 (3.90) & 18.36 (5.42) & 0.77 (0.05) \\
       & RNP-Pool & 18.00 (0.00) & 0.00 (0.00) & 17.85 (0.67) & 0.15 (0.67) & 2.62 (0.80) & 0.87 (0.02) \\
        \specialrule{0em}{3pt}{3pt}
       Error 3 
       & RNP-Int & 17.20 (1.51) & 0.00 (0.00)  & 15.85 (2.56) & 1.40 (3.20) & 5.82 (5.46) & 0.80 (0.03) \\
       Cauchy(0,1) & NRNP-Int & 16.30 (1.87) & 0.00 (0.00) & 3.25 (3.68) & 21.80 (3.38) & 234.62 (41.43) & 0.52 (0.09)\\
       & RP-Int   & 17.60 (0.82) & 0.00 (0.00) & 11.45 (1.10) & 1.15 (1.76) & 7.61 (1.99) & 0.78 (0.02) \\
       & NRP-Int & 16.40 (1.90) & 0.00 (0.00) & 2.05 (2.31) & 22.20 (3.38) & 302.4 (62.04) & 0.49 (0.10)\\
       & RNP-Meta  & 5.85 (1.64) & 0.00 (0.00) & 8.00 (2.13) & 19.75 (3.01) & 27.79 (4.40) & 0.66 (0.05) \\
       & RNP-Pool & 18.00 (0.00) & 0.00 (0.00) & 15.90 (2.40) & 1.35 (3.57) & 5.61 (1.02) & 0.79 (0.03)\\
 \bottomrule
   \end{tabular}
   \label{tab:AFTsce1-2}
\end{table}

\begin{table}[htbp]
   \centering
   \caption{
   Simulation Scenarios 3-4 for data with a right-censored survival outcome. In each cell, mean (standard deviation).} 
   \begin{tabular}{llcccccc} 
      \toprule
      & & \multicolumn{2}{c}{{\bf Commonality identification}} & \multicolumn{2}{c}{{\bf Variable selection}}  \\
      \cmidrule(r){3-4} \cmidrule(l){5-6} 
      & {\bf Methods} & \bf {TP-ind} &{\bf FP-ind }&  {\bf TP-var}  & {\bf FP-var} & {\bf RMISE} & {\bf {\color{black} Cstat}} \\
      \midrule
       & & & & \bf{ Scenario 3} \\
       Error 1 
       & RNP-Int & 0.00 (0.00) & 3.50 (2.35) & 12.00 (1.92) & 0.95 (1.19) & 7.40 (3.41) & 0.90 (0.03) \\
     N(0,1)  & NRNP-Int & 0.00 (0.00) & 1.50 (0.95) & 13.35 (0.67) & 2.10 (1.94) & 5.66 (1.43) & 0.92 (0.01)  \\
       & RP-Int  &0.00 (0.00) & 7.65 (1.73) & 9.55 (1.00) & 1.50 (1.40) & 9.22 (2.24) & 0.86 (0.02)  \\
       & NRP-Int  &0.00 (0.00) & 6.65 (1.46) & 10.00 (0.65) & 12.45 (3.47) & 15.88 (2.76) & 0.83 (0.02)\\
       & RNP-Meta  & 0.00 (0.00) & 0.45 (1.10) & 12.25 (1.52) & 9.20 (4.12) & 9.82 (5.22) & 0.88  (0.04) \\
       & RNP-Pool & 0.00 (0.00) & 18.00 (0.00) & 11.25 (2.40) & 2.85 (0.99) & 19.40 (1.39) & 0.72 (0.04) \\
       \specialrule{0em}{3pt}{3pt}
       Error 2 
       & RNP-Int & 0.00 (0.00) & 4.60 (2.26) & 11.90 (1.17) & 1.30 (1.59) & 8.41 (3.80) & 0.85 (0.03)\\
       70\% N(0,1)& NRNP-Int & 0.00 (0.00) & 9.35 (4.58) & 7.75 (3.61) & 16.70 (5.66) & 52.78 (9.62) & 0.71 (0.11) \\
     +30\%Cauchy(0,1)   & RP-Int   & 0.00 (0.00) & 7.50 (1.28) & 9.70 (0.47) & 1.05 (0.83) & 8.76 (1.22) & 0.84 (0.02)\\
       & NRP-Int & 0.00 (0.00) & 12.10 (3.39) & 5.30 (3.26) & 19.85 (4.65) & 63.82 (11.16) & 0.67 (0.11)\\
       & RNP-Meta  & 0.00 (0.00) & 2.65 (1.60) & 10.85 (1.87) & 12.85 (4.25) & 14.61 (5.12) & 0.81 (0.04) \\
       & RNP-Pool & 0.00 (0.00) & 18.00 (0.00) & 11.00 (1.89) & 2.80 (1.32) & 19.37 (2.17) & 0.70 (0.04)  \\
       \specialrule{0em}{3pt}{3pt}
       Error 3 
       & RNP-Int & 0.00 (0.00) & 6.20 (2.69) & 11.10 (1.48) & 2.60 (2.21) & 10.50 (4.39) & 0.79 (0.04) \\
       Cauchy(0,1) & NRNP-Int & 0.00 (0.00) & 15.40 (2.98) & 2.30 (2.54) & 22.00 (3.34) & 247.11 (48.35) & 0.51 (0.09) \\
       & RP-Int   & 0.00 (0.00) & 9.50 (1.88) & 8.60 (1.35) & 2.35 (2.46) & 11.82 (4.01) & 0.77 (0.04) \\ 
       & NRP-Int &  0.00 (0.00) & 15.35 (2.48) & 2.10 (2.53) & 21.55 (3.15) & 319.76 (77.76) & 0.49 (0.10)\\
       & RNP-Meta  &  0.00 (0.00) & 5.15 (2.76) & 8.45 (2.26) & 19.20 (3.83) & 24.58 (7.61) & 0.72 (0.06)  \\
       & RNP-Pool & 0.00 (0.00) & 18.00 (0.00) & 10.50 (1.57) & 4.20 (2.67) & 20.61 (3.03) & 0.67 (0.03) \\
       \midrule
        & & &  & \bf{Scenario 4} \\
        Error 1
       & RNP-Int & 6.45 (1.28) & 1.75 (1.20) & 15.35 (1.81) & 1.40 (1.35) & 8.76 (3.37) & 0.91 (0.03)  \\
      N(0,1)  & NRNP-Int & 6.60 (1.10) & 1.45 (0.89) & 17.85 (0.49) & 2.75 (2.77) & 6.03 (2.82) & 0.93 (0.02)  \\
       & RP-Int & 7.00 (0.97) & 1.90 (2.15) & 17.35 (1.09) & 1.30 (1.30) & 3.62 (2.20) & 0.94 (0.01)\\
       & NRP-Int & 6.85 (1.04) & 0.20 (0.62) & 18.00 (0.00) & 0.85 (1.18) & 2.01 (0.83) & 0.95 (0.01) \\
       & RNP-Meta  & 2.25 (0.64) & 2.65 (2.28) & 14.70 (2.36) & 10.55 (4.69) & 13.44 (6.42) & 0.89 (0.04) \\
       & RNP-Pool &  8.00 (0.00) & 19.00 (0.00) & 12.05 (2.19) & 5.20 (1.54) & 26.62 (2.79) & 0.68 (0.03)\\
        \specialrule{0em}{3pt}{3pt}
       Error 2 
       & RNP-Int &  7.10 (0.85) & 4.15 (2.57) & 13.40 (3.35) & 2.60 (1.57) & 11.38 (4.81) & 0.86 (0.04)\\
       70\%$N(0,1)$ 
       & NRNP-Int  & 6.55 (1.50) & 8.70 (4.33) & 10.10 (4.17) & 26.30 (7.68) & 63.17 (10.56) & 0.74 (0.08)\\
       +30\% Cauchy(0,1)  & RP-Int &  6.65 (1.23) & 2.70 (2.45) & 16.55 (1.54) & 1.60 (1.27) & 4.59 (2.20) & 0.90 (0.02)\\
       & NRP-Int & 6.50 (1.32) & 8.45 (5.50) & 10.15 (5.29) & 25.50 (9.16) &  70.83 (14.19) & 0.75 (0.10)\\
       & RNP-Meta  & 3.55 (1.76) & 3.60 (2.06) & 12.75 (2.20) & 13.45 (3.86) & 16.80 (4.79) & 0.85 (0.03) \\
       & RNP-Pool & 8.00 (0.00) & 19.00 (0.00) & 11.00 (2.32) & 5.35 (1.53) & 27.30 (1.82) & 0.66 (0.03) \\
        \specialrule{0em}{3pt}{3pt}
       Error 3
       & RNP-Int & 7.10 (1.12) & 5.20 (2.61) & 10.00 (2.51) & 4.90 (2.13) & 17.39 (4.22) & 0.79 (0.04) \\
      Cauchy(0,1) & NRNP-Int & 7.20 (0.89) & 15.30 (2.49) & 3.00 (2.47) & 34.30 (3.76) & 277.58 (45.76) & 0.54 (0.08) \\
       & RP-Int   & 6.75 (1.29) & 4.75 (2.12) & 14.30 (1.78) & 2.05 (1.67) & 7.62 (2.58) & 0.83 (0.03) \\
       & NRP-Int & 6.75 (1.29) & 14.40 (2.74) & 3.75 (2.90) & 32.05 (5.64) & 373.4 (74.28) & 0.53 (0.12) \\
       & RNP-Meta  & 4.70 (1.26) & 7.30 (2.08) & 9.25 (1.59) & 18.45 (2.74) & 26.22 (6.00) & 0.75 (0.04) \\
       & RNP-Pool & 8.00 (0.00) & 19.00 (0.00) & 8.10 (2.45) & 6.60 (2.06) & 31.19 (4.82) & 0.63 (0.05)\\
 \bottomrule 
   \end{tabular}
   \label{tab:AFTsce3-4}
\end{table}

\begin{table}[htbp]
   \centering
   \caption{
   {\color{black} Simulation Scenarios 1 for data with a continuous outcome under unbalanced subgroups. In each cell, mean (standard deviation).}} 
   \begin{tabular}{llcccccc} 
      \toprule
      & & \multicolumn{2}{c}{{\bf Commonality identification}} & \multicolumn{2}{c}{{\bf Variable selection}}  \\
      \cmidrule(r){3-4} \cmidrule(l){5-6} 
      & {\bf Methods} & \bf {TP-ind} &{\bf FP-ind }&  {\bf TP-var}  & {\bf FP-var} & {\bf RMISE} & {\bf MAE} \\
      \midrule
       Error 1 
       & RNP-Int & 0.00 (0.00) & 3.50 (2.35) & 12.00 (1.92) & 0.95 (1.19) & 7.40 (3.41) & 4.42 (0.81) \\
     N(0,1)  & NRNP-Int & 0.00 (0.00) & 1.50 (0.95) & 13.35 (0.67) & 2.10 (1.94) & 5.66 (1.43) & 3.80 (0.55)  \\
       & RP-Int  &0.00 (0.00) & 7.65 (1.73) & 9.55 (1.00) & 1.50 (1.40) & 9.22 (2.24) & 5.98 (0.85)  \\
       & NRP-Int  &0.00 (0.00) & 6.65 (1.46) & 10.00 (0.65) & 12.45 (3.47) & 15.88 (2.76) & 7.05 (0.98)\\
       & RNP-Meta  & 0.00 (0.00) & 0.45 (1.10) & 12.25 (1.52) & 9.20 (4.12) & 9.82 (5.22) &5.08  (1.88) \\
       & RNP-Pool & 0.00 (0.00) & 18.00 (0.00) & 11.25 (2.40) & 2.85 (0.99) & 19.40 (1.39) & 10.92 (1.91) \\
       \specialrule{0em}{3pt}{3pt}
       Error 2 
       & RNP-Int & 0.00 (0.00) & 4.60 (2.26) & 11.90 (1.17) & 1.30 (1.59) & 8.41 (3.80) & 15.87 (6.28)\\
       70\% N(0,1)& NRNP-Int & 0.00 (0.00) & 9.35 (4.58) & 7.75 (3.61) & 16.70 (5.66) & 52.78 (9.62) & 27.87 (6.33) \\
     +30\%Cauchy(0,1)   & RP-Int   & 0.00 (0.00) & 7.50 (1.28) & 9.70 (0.47) & 1.05 (0.83) & 8.76 (1.22) & 16.76 (6.31)\\
       & NRP-Int & 0.00 (0.00) & 12.10 (3.39) & 5.30 (3.26) & 19.85 (4.65) & 63.82 (11.16) & 29.33 (6.38)\\
       & RNP-Meta  & 0.00 (0.00) & 2.65 (1.60) & 10.85 (1.87) & 12.85 (4.25) & 14.61 (5.12) & 17.39 (6.37) \\
       & RNP-Pool & 0.00 (0.00) & 18.00 (0.00) & 11.00 (1.89) & 2.80 (1.32) & 19.37 (2.17) & 21.70 (6.33)  \\
       \specialrule{0em}{3pt}{3pt}
       Error 3 
       & RNP-Int & 0.00 (0.00) & 6.20 (2.69) & 11.10 (1.48) & 2.60 (2.21) & 10.50 (4.39) & 19.31 (1.94) \\
       Cauchy(0,1) & NRNP-Int & 0.00 (0.00) & 15.40 (2.98) & 2.30 (2.54) & 22.00 (3.34) & 247.11 (48.35) & 98.75 (19.37) \\
       & RP-Int   & 0.00 (0.00) & 9.50 (1.88) & 8.60 (1.35) & 2.35 (2.46) & 11.82 (4.01) & 20.25 (2.00) \\ 
       & NRP-Int &  0.00 (0.00) & 15.35 (2.48) & 2.10 (2.53) & 21.55 (3.15) & 319.76 (77.76) & 101.11 (19.22)\\
       & RNP-Meta  &  0.00 (0.00) & 5.15 (2.76) & 8.45 (2.26) & 19.20 (3.83) & 24.58 (7.61) & 22.40 (2.13)  \\
       & RNP-Pool & 0.00 (0.00) & 18.00 (0.00) & 10.50 (1.57) & 4.20 (2.67) & 20.61 (3.03) & 23.99 (2.00) \\
       \bottomrule  
    \end{tabular}
   \label{tab:unbalance1}
\end{table}

\begin{table}[htbp]
   \centering
   \caption{
   {\color{black} Simulation Scenarios 1 for data with a right-censored survival outcome under unbalanced subgroups. In each cell, mean (standard deviation).}} 
   \begin{tabular}{llcccccc} 
      \toprule
      & & \multicolumn{2}{c}{{\bf Commonality identification}} & \multicolumn{2}{c}{{\bf Variable selection}}  \\
      \cmidrule(r){3-4} \cmidrule(l){5-6} 
      & {\bf Methods} & \bf {TP-ind} &{\bf FP-ind }&  {\bf TP-var}  & {\bf FP-var} & {\bf RMISE} & {\bf Cstat} \\
      \midrule
        Error 1
       & RNP-Int & 5.80 (1.36) & 2.05 (1.15) & 10.95 (0.61) & 0.9 (1.33) & 3.40 (0.98) & 0.89 (0.02)  \\
       N(0,1)  & NRNP-Int & 6.25 (1.12) & 0.35 (0.75) & 11.80 (0.41) & 2.21 (1.32) & 3.03 (1.20) & 0.90 (0.01)  \\
       & RP-Int & 6.40 (0.75) & 4.40 (0.82) & 8.60 (0.82) & 1.05 (1.15) & 4.78 (1.59) & 0.85 (0.02)\\
       & NRP-Int & 5.55 (1.23) & 3.80 (0.77) & 9.35 (0.99) & 11.50 (3.53) & 9.21 (2.01) & 0.83 (0.02) \\
       & RNP-Meta  & 1.00 (0.32) & 1.60 (1.19) & 9.65 (1.09) & 8.75 (2.17) & 8.81 (2.44) & 0.86 (0.02) \\
       & RNP-Pool &  7.00 (0.00) & 11.00 (0.00) & 9.40 (0.68) & 3.65 (1.14) & 10.80 (0.61) & 0.75 (0.03)\\
        \specialrule{0em}{3pt}{3pt}
       Error 2 
       & RNP-Int &  6.20 (1.20) & 2.50 (1.24) & 10.60 (0.88) & 1.10 (1.29) & 4.00 (2.41) & 0.85 (0.03)\\
       70\%$N(0,1)$ 
       & NRNP-Int  & 5.75 (1.21) & 5.35 (3.17) & 6.70 (3.54) & 18.10 (4.39) & 46.21 (8.96) & 0.65 (0.13)\\
       +30\% Cauchy(0,1)  & RP-Int &  6.10 (0.85) & 4.55 (1.19) & 8.60 (0.82) & 1.20 (1.06) & 5.22 (1.59) & 0.82 (0.02)\\
       & NRP-Int & 6.25 (0.85) & 6.80 (2.59) & 5.25 (3.28) & 19.85 (3.23) &  54.40 (10.72) & 0.64 (0.13)\\
       & RNP-Meta  & 1.10 (0.45) & 2.80 (1.70) & 8.90 (1.68) & 11.60 (2.80) & 11.60 (3.38) & 0.80 (0.04) \\
       & RNP-Pool & 7.00 (0.00) & 11.00 (0.00) & 9.05 (0.22) & 3.40 (1.00) & 11.00 (1.00) & 0.73 (0.03) \\
        \specialrule{0em}{3pt}{3pt}
       Error 3
       & RNP-Int & 6.25 (1.45) & 3.65 (1.42) & 9.40 (1.96) & 2.70 (1.72) & 7.60 (3.40) & 0.77 (0.03) \\
      Cauchy(0,1) & NRNP-Int & 6.40 (0.75) & 9.20 (1.96) & 1.75 (2.36) & 22.40 (2.50) & 232.22 (40.43) & 0.50 (0.08) \\
       & RP-Int   & 6.30 (0.98) & 4.65 (0.99) & 8.25 (1.18) & 1.75 (1.37) & 7.40 (1.82) & 0.75 (0.03) \\
       & NRP-Int & 6.10 (0.97) & 9.70 (1.46) & 1.40 (1.67) & 22.45 (3.50) & 300.78 (55.32) & 0.50 (0.09) \\
       & RNP-Meta  & 1.45 (0.69) & 3.80 (1.28) & 7.35 (1.53) & 16.05 (3.02) & 17.89 (3.22) & 0.71 (0.04) \\
       & RNP-Pool & 7.00 (0.00) & 11.00 (0.00) & 8.95 (1.36) & 3.50 (0.89) & 12.20 (1.19) & 0.69 (0.03)\\
 \bottomrule 
   \end{tabular}
   \label{tab:unbalance2}
\end{table}



\clearpage

\end{document}